\documentclass[final,3p,times]{elsarticle}

\usepackage{amssymb}
\usepackage{amsmath}
\usepackage{tikz}
\usepackage{algorithm2e}
\usepackage[
    colorlinks,
    linkcolor={blue}
    citecolor={blue},
    urlcolor={blue}]
    {hyperref}
\usepackage{subcaption}
\usepackage{tabularx}
\usepackage{enumitem}

\usepackage{listings}
\newcounter{bla}

\newcommand{\dcaplus}{DCA$^+\:$}
\newcommand{\bk}{{\bf k}}
\newcommand{\bK}{{\bf K}}
\newcommand{\br}{{\bf r}}
\newcommand{\bq}{{\bf q}}
\newcommand{\wn}{{i\omega_n}}
\newcommand{\vm}{{i\nu_m}}
\newcommand{\Gb}{\bar{G}}

\journal{Computer Physics Communications}

\begin{document}

\begin{frontmatter}



\title{DCA++: A software framework to solve correlated electron problems with modern quantum cluster methods}


\author[a]{Urs R. H\"ahner\corref{author}}
\author[b]{Gonzalo Alvarez}
\author[b]{Thomas A. Maier}
\author[c]{Raffaele Solc\`a}
\author[d]{Peter Staar}
\author[b]{Michael S. Summers}
\author[a,b,c]{Thomas C. Schulthess}

\cortext[author] {Corresponding author.\\\textit{E-mail address:} haehneru@itp.phys.ethz.ch}
\address[a]{Institute for Theoretical Physics, ETH Zurich, 8093 Zurich, Switzerland}
\address[b]{Computational Science and Engineering Division, Center for Nanophase Materials Sciences, Oak Ridge National Laboratory, Oak Ridge, Tennessee 37831, USA}
\address[c]{Swiss National Supercomputing Centre, ETH Zurich, 6900 Lugano, Switzerland}
\address[d]{IBM Research -- Zurich, 8803 R\"uschlikon, Switzerland}

\begin{abstract}
We present the first open release of the DCA++ project, a high-performance research software framework to solve quantum many-body problems with cutting edge quantum cluster algorithms.
DCA++ implements the dynamical cluster approximation~(DCA) and its DCA$^+$ extension with a continuous self-energy.
The algorithms capture nonlocal correlations in strongly correlated electron systems, thereby giving insight into high-$T_c$ superconductivity.
The code's scalability allows efficient usage of systems at all scales, from workstations to leadership computers.
With regard to the increasing heterogeneity of modern computing machines, DCA++ provides portable performance on conventional and emerging new architectures, such as hybrid CPU-GPU, sustaining multiple petaflops on ORNL's Titan and CSCS' Piz Daint supercomputers.
Moreover, we show how sustainable and scalable development of the code base has been achieved by adopting standard techniques of the software industry.
These include employing a distributed version control system, applying test-driven development and following continuous integration. \\
\end{abstract}

\begin{keyword}
Strongly correlated electron systems;
Quantum cluster algorithms;
Dynamical cluster approximation;
Continuous-time quantum Monte Carlo;
Extreme-scale computing;
Sustainable software development
\end{keyword}

\end{frontmatter}



{\bf PROGRAM SUMMARY}

\begin{small}
\noindent
{\em Program Title:} DCA++ \\
{\em Licensing provisions:} BSD-3-Clause \\
{\em Programming language:} C++14 and CUDA \\
{\em Nature of problem:}
Understanding the fascinating physics of strongly correlated electron systems requires the development of sophisticated algorithms and their implementation on leadership computing systems. \\
{\em Solution method:}
The DCA++ code provides a highly scalable and efficient implementation of the dynamical cluster approximation~(DCA) and its DCA$^+$ extension. \\
\end{small}

\section{Introduction}
\label{sec:introduction}

Computer simulations have become indispensable across all areas of science and physics in particular.
In condensed matter physics significant computational effort is devoted to solving the quantum many-body problem of interacting electrons.
Strong correlations between electrons are the origin of a variety of fascinating phenomena, most prominently high-temperature superconductivity.
The enormous complexity of the quantum many-body problem - the Hilbert space grows exponentially with the system size - and lack of approximations and simplifications in the physically relevant regime pose a great challenge to date.

Progress in solving the many-electron problem has been made by introducing models that comprise the fundamental mechanisms that lead to the observed phenomena.
The two-dimensional Hubbard model, for instance, is believed to capture the important physics in the superconducting cuprates~\cite{Anderson:1987, Zhang:1988jf}.
The Hubbard Hamiltonian describes fermions on a lattice, where the particles are allowed to hop between lattice sites and interact through Coulomb repulsion when they occupy the same site.
Despite the simple structure, there exists no exact solution except for the one-dimensional case.

To get insight into the physics described by the Hubbard model, one has to resort to numerical methods.
Exact diagonalization~(ED) and a variety of quantum Monte Carlo~(QMC) algorithms solve the model on a finite-size lattice.
While ED is restricted to small systems due to the exponential scaling of the problem size with the number of lattice sites, the negative sign problem prevents QMC calculations on large lattices or at low temperatures.
Finite-size effects arise from the truncation of the infinite lattice to a finite number of sites.

Mean-field theories choose a different approach and are formulated in the thermodynamic limit, that is on the infinite lattice.
In dynamical mean-field theory (DMFT)~\cite{Georges:1996hv} the infinite lattice problem becomes tractable by reducing it to the self-consistent solution of an effective impurity model.
Underlying the single-site approximation is the assumption that the self-energy is local in real space, which holds in the limit of infinite dimensions.
However, spatially nonlocal correlations are often essential to describe phase transitions, which is the reason why DMFT has been extended by a number of quantum cluster algorithms~\cite{Maier:2005do}.
The dynamical cluster approximation~(DCA)~\cite{Maier:2005do, Hettler:2000es, Jarrell:2001jj} maps the bulk lattice problem to a finite-size periodic cluster that is self-consistently embedded in a dynamical mean-field.
The effective cluster problem is best solved by means of continuous-time QMC methods~\cite{Gull:2011jd}.
Compared to finite-size QMC calculations, DCA's mean-field approach not only reduces the fermion sign problem~\cite{Jarrell:2001jj}, but also generally leads to results that converge faster to the thermodynamic limit~\cite{Jarrell:2001jj, Moukouri:2001ch}.

The DCA++ project provides an efficient and highly scalable C++ implementation of the DCA algorithm and the DCA$^{+}$ extension that introduces a continuous lattice self-energy \cite{Staar:2013ec}.
The code has been developed by a team at ORNL and ETH Zurich, which, besides modern software design, has the focus on developing and implementing sophisticated algorithms~\cite{Nukala:2009gh, Staar:2012fj} and exploiting the benefits of emerging architectures, such as hybrid CPU-GPU systems.
In addition to setting benchmarks for extreme scale scientific applications~\cite{Alvarez:2008wj, Staar:2013ik}, the DCA++ code has provided the numerical evidence for a series of scientific publications in the field of strongly correlated electron systems.
Studies of the pairing mechanism in unconventional superconductors~\cite{Staar:2016fl} and the influence of near neighbor Coulomb repulsion on $d$-wave superconductivity~\cite{Jiang:2018jk}, for example, were based on DCA calculations with DCA++.

In software-driven research, scientific productivity is strongly coupled to software productivity.
Hence, the scientific output of a research group can be held by challenges such as new computer architectures, advanced algorithms or changing teams of developers.
To prevent this productivity collapse, software development needs to be sustainable and scalable by producing comprehensible, maintainable, and extensible code.
At the same time, it is essential to release changes to the code base, from new features to bug fixes, rapidly to the users, an ability that usually falls under the term \emph{continuous delivery}.
Last but not least, the scientific standard demands correctness, credibility and reproducibility of numerical results in published work.
To fulfill these requirements in a challenging environment formed by complex algorithms, performance sensitive codes, the diversity of architectures, and the multidisciplinary of teams, the DCA++ project employs well-proven tools and successful techniques of the software industry~\cite{Haehner:2018dca-development}.
While adopting these methods can require an effort, we believe that they represent a substantial factor for a research code to become a long-lived software project.

This paper accompanies the first open release of the DCA++ code and is structured as follows:
Section~\ref{sec:methods} reviews the underlying DCA and \dcaplus algorithms, briefly discusses the continuous-time auxiliary-field and continuous-time hybridization expansion QMC algorithms, and outlines the computation of two-particle correlation functions.
Section~\ref{sec:code} gives an overview of the main components of the DCA++ code, addresses prerequisites and the CMake-based building routine, and explains how to run a DCA or \dcaplus calculation.
Section~\ref{sec:development} explores the software development practice that is followed in the DCA++ project, but whose applicability extends to many other collaborative scientific software efforts.
Section~\ref{sec:examples} provides two typical use cases supplemented by performance analyses.
The paper concludes with an outline of future development plans and a summary of ways to contribute back to the DCA++ project.

\section{Formalism and methods}
\label{sec:methods}

\subsection{The dynamical cluster approximation}
\label{subsec:dca}

\begin{figure}[h]
    \centering
    \includegraphics[scale=0.3]{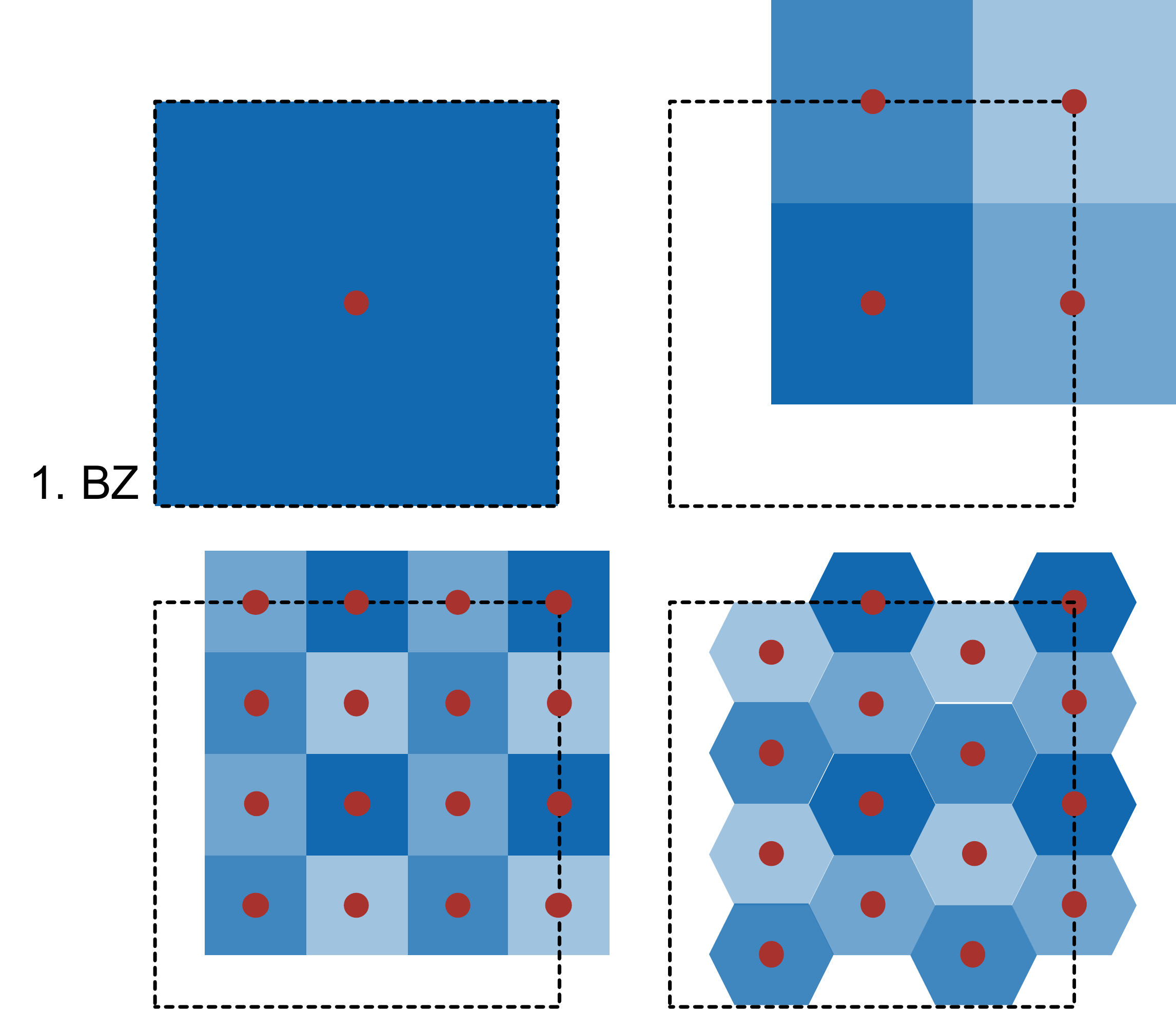}
    \caption{
        (Color online)
        Cluster momenta $\bK$ (red dots) and standard coarse-graining patches for different cluster sizes and shapes:
        $N_c=1$ (DMFT) (top left), $N_c=4$ (top right), $N_c=16A$ (bottom left), and $N_c=16B$ (bottom right).
    }
    \label{fig:patches}
\end{figure}

Like finite-size methods, the DCA algorithm replaces the infinite real-space lattice by a finite-size cluster to reduce the complexity of the problem.
The reduction of the infinite lattice to a cluster of $N_c$ sites corresponds to a discretization of the first Brillouin zone into a set of $N_c$ cluster momenta $\bK$  (see Fig.~\ref{fig:patches}).
In contrast to finite-size methods, which solve the cluster in isolation, the DCA embeds the cluster in a mean field to retain information about the degrees of freedom of the lattice that are not contained in the cluster.
This is achieved through a coarse-graining procedure in momentum space, where these degrees of freedom are averaged out.
For this purpose, the first Brillouin zone~(BZ) is divided into $N_c$ patches (see Fig.~\ref{fig:patches}).
Each patch is centered around a unique cluster momentum $\bK$ and represented by a patch function $\phi_\bK(\bk)$, the characteristic function of the patch,
\begin{equation}
    \phi_\bK(\bk) = \left\{\begin{array}{ll}
        1, & \bk \in \bK^\mathrm{th} \, \mathrm{patch}, \\
        0, & \mathrm{otherwise}.
    \end{array}\right.
\end{equation}
In addition to the orthogonality condition,
\begin{equation}
\label{eq:orthogonality_condition}
    \frac{N_c}{V_\mathrm{BZ}} \int_\mathrm{BZ} \! d\bk \, \phi_\bK(\bk) \phi_{\bK'}(\bk) = \delta_{\bK, \bK'} \,,
\end{equation}
we require the patches to have equal size and shape,
\begin{equation}
\label{eq:patch_translation_condition}
    \phi_\bK(\bk) = \phi(\bk-\bK) \,,
\end{equation}
and to exhibit inversion symmetry,
\begin{equation}
\label{eq:patch_inversion_symmetry}
    \phi(\bk-\bK) = \phi(\bK-\bk)\,.
\end{equation}
Despite these restrictions, the choice of patches is far from unique and it is even possible to create interleaved shapes~\cite{Staar:2016ia}.
In the standard coarse-graining the patches are defined as the Brillouin zones of the superlattice as illustrated in Fig.~\ref{fig:patches}.

\begin{figure}
    \centering
    \includegraphics[scale=0.45]{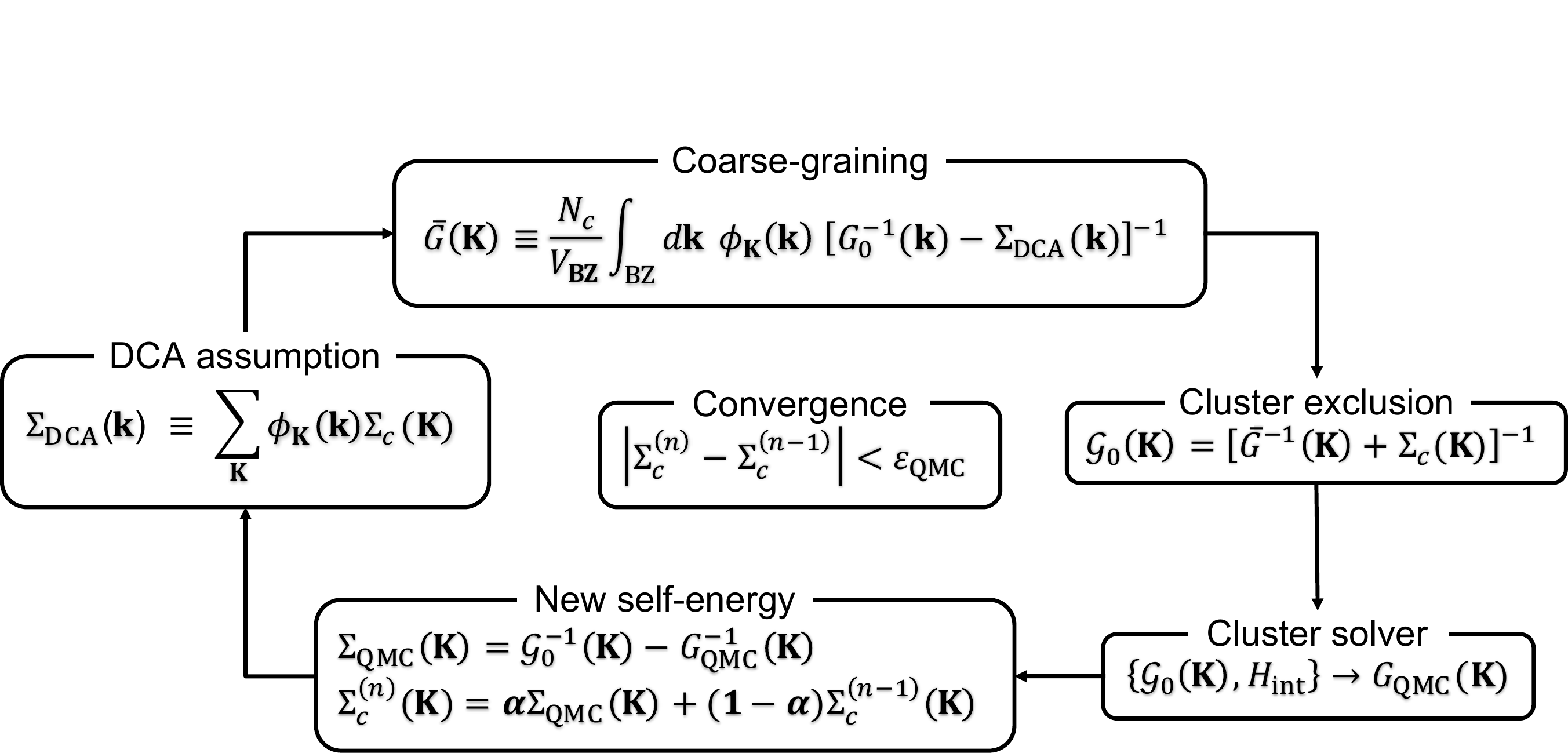}
    \caption{
        The DCA self-consistency loop:
        In accordance with the DCA assumption, the lattice self-energy $\Sigma_\mathrm{DCA}(\bk)$ is constructed as a piecewise constant continuation of the cluster self-energy $\Sigma_c(\bK)$.
        The former is then used to calculate the coarse-grained Green's function $\Gb(\bK)$.
        Given $\Gb(\bK)$ and $\Sigma_c(\bK)$, the bare cluster Green's function $\mathcal{G}_0(\bK)$ is obtained from the Dyson equation.
        $\mathcal{G}_0(\bK)$ complemented by the interaction Hamiltonian $H_\mathrm{int}$ defines the effective cluster problem, which is solved, in general, by means of QMC techniques.
        The self-consistency loop is closed by producing a new cluster self-energy $\Sigma_c(\bK)$.
    }
    \label{fig:dca}
\end{figure}

The underlying assumption of the DCA is that the lattice self-energy $\Sigma(\bk, \wn)$ is a slowly varying function in momentum space and in the vicinity of a cluster momentum $\bK$ well approximated by the self-energy of the cluster $\Sigma_c(\bK, \wn)$,
\begin{equation}
    \Sigma(\bk=\bK+\tilde{\bk}, \wn) \approx \Sigma_c(\bK, \wn) \,.
\end{equation}
Based on this assumption, we can approximate the lattice self-energy $\Sigma(\bk, \wn)$ by a piecewise constant continuation of the cluster self-energy $\Sigma_c(\bK, \wn)$ by means of the patch functions $\phi_\bK(\bk)$,
\begin{equation}
\label{eq:sigma_dca}
    \Sigma_\mathrm{DCA}(\bk, \wn) \equiv \sum_\bK \phi_\bK(\bk) \, \Sigma_c(\bK, \wn) \,.
\end{equation}
Unlike finite-size methods, the DCA, using above approximation of the lattice self-energy, replaces the Green's function of the cluster $G_c(\bK, \wn)$ with the coarse-grained average of the lattice Green's function,
\begin{equation}
\label{eq:G_coarsegrained}
    G_c(\bK, \wn) \rightarrow \Gb(\bK, \wn) = \frac{N_c}{V_\mathrm{BZ}} \int_\mathrm{BZ} \! d\bk \, \phi_\bK(\bk) \left[ G_0^{-1}(\bk, \wn) - \Sigma_\mathrm{DCA}(\bk, \wn) \right]^{-1} \,.
\end{equation}
$G_0(\bk, \wn)$ is the non-interacting lattice Green's function given by
\begin{equation}
    G_0(\bk, \wn) = \left[ \wn - \varepsilon_\bk + \mu \right]^{-1} \,.
\end{equation}
To formulate the effective cluster problem of the DCA, we need the corresponding bare propagator of the cluster $\mathcal{G}_0(\bK, \wn)$.
We obtain it by reversing the Dyson equation in the \emph{cluster-exclusion step},
\begin{equation}
\label{eq:cluster-exclusion}
    \mathcal{G}_0(\bK, \wn) = \left[ \Gb^{-1}(\bK, \wn) + \Sigma_c(\bK, \wn) \right]^{-1} \,.
\end{equation}
$\mathcal{G}_0(\bK, \wn)$ is fundamentally different from the free lattice propagator evaluated at the cluster momenta $\bK$, $G_0(\bK, \wn)$.
It incorporates the coupling of the cluster to the mean-field that represents the remaining degrees of freedom of the lattice.
The effective cluster problem is completed by the interacting part of the Hamiltonian $H_\mathrm{int}$ and best solved by QMC methods to calculate a new cluster self-energy $\Sigma_c(\bK, \wn)$,
\begin{equation}
\label{eq:QMC}
    \left\{ \mathcal{G}_0(\bK, \wn), H_\mathrm{int} \right\} \stackrel{\mathrm{QMC}}{\longrightarrow} \Sigma_c(\bK, \wn) \,.
\end{equation}
This closes the DCA self-consistency loop.

To start the loop, one has to provide an initial cluster self-energy $\Sigma_c(\bK, \wn)$, which in practice is either zero or the result of a previous calculation for a slightly higher temperature (see Section~\ref{subsec:running}).
Steps (\ref{eq:sigma_dca}), (\ref{eq:G_coarsegrained}), (\ref{eq:cluster-exclusion}) and (\ref{eq:QMC}) are then iterated until convergence is reached, where the criterion for convergence is dictated by the stochastic error of the QMC solver $\varepsilon_\mathrm{QMC}$,
\begin{equation}
    |\Sigma_c^{(n)} - \Sigma_c^{(n-1)}| < \varepsilon_\mathrm{QMC} \,.
\end{equation}
To stabilize convergence of the loop, the self-energy produced by the QMC solver $\Sigma_\mathrm{QMC}(\bK, \wn)$ is often mixed with the previous value of the cluster self-energy $\Sigma_c(\bK, \wn)$,
\begin{equation}
\label{eq:self-energy-mixing}
    \Sigma_c^{(n)} = \alpha \, \Sigma_\mathrm{QMC} + (1-\alpha) \, \Sigma_c^{(n-1)} \,, \quad 0<\alpha\leq1 \,.
\end{equation}
The complete DCA algorithm is summarized in Fig.~\ref{fig:dca}.

Relying on QMC methods, the DCA algorithm inherits their negative sign problem.
Nevertheless, the sign problem turns out to be less severe than in finite-size lattice QMC simulations, an effect the dynamical mean-field has been credited with~\cite{Jarrell:2001jj}.

Like finite-size methods, the DCA yield the exact result of the infinite lattice problem as $N_c \rightarrow \infty$.
In other words, if clusters large enough to extrapolate to infinite cluster size are accessible, we can obtain exact results for the thermodynamic limit by finite-size scaling.
In doing so, convergence with respect to cluster size indicates that the current size of the cluster captures all relevant nonlocal correlations.
In the opposite limit, i.e. for $N_c = 1$, the DCA recovers DMFT.

\subsection{The \dcaplus algorithm}
\label{subsec:dca_plus}

\begin{figure}[h!]
    \centering
    \includegraphics[width=\columnwidth]{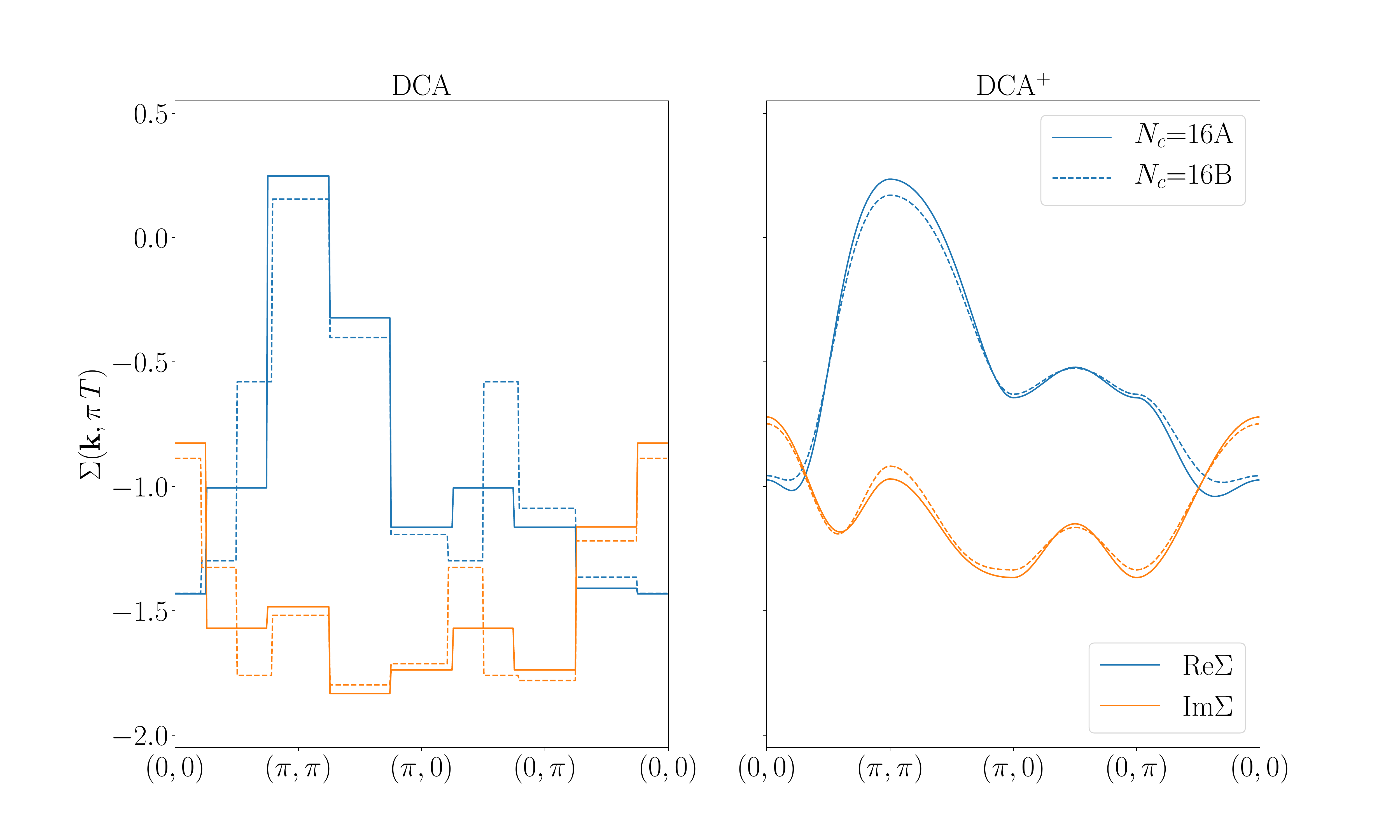}
    \caption{
        (Color online)
        Momentum dependence of the DCA (left) and DCA$^+$ (right) lattice self-energy at the lowest Matsubara frequency for a 2D Hubbard model, Eq.~(\ref{eq:Hubbard}), with $U=7t$ and $\langle n \rangle = 0.95$ at a temperature $T=0.2t$.
        Results are shown for the two different 16-site clusters of Fig.~\ref{fig:patches}.}
    \label{fig:DCA_vs_DCA+_Sigma-band-structure}
\end{figure}

The cluster shape dependence represents one of the major drawbacks of the DCA algorithm (see Fig.~\ref{fig:DCA_vs_DCA+_Sigma-band-structure}).
The expansion of the lattice self-energy $\Sigma_\mathrm{DCA}(\bk, \wn)$, Eq.~(\ref{eq:sigma_dca}), evidently depends on the shape of the coarse-graining patches and is therefore sensitive to the definition of the cluster.
A way to cure this problem is to abandon the representation of the lattice self-energy in terms of step functions and find a continuous description.
Simple interpolation schemes of the cluster self-energy, however, have been shown to lead to causality violations~\cite{Hettler:2000es}.
Instead, we start from a coarse-graining equation for the self-energy, equivalent to the expression for the Green's function in Eq.~(\ref{eq:G_coarsegrained}).
Inverting Eq.~(\ref{eq:sigma_dca}) with the help of the orthogonality condition of the patch functions, Eq.~(\ref{eq:orthogonality_condition}), we obtain the relation,
\begin{equation}
    \label{eq:coarse-graining_condition_self-energy}
    \Sigma_c(\bK, \wn) = \frac{N_c}{V_\mathrm{BZ}} \int_\mathrm{BZ} \! d\bk \, \phi_\bK(\bk) \, \Sigma_{\mathrm{DCA}^+}(\bk, \wn) \,.
\end{equation}
This equation implies that we need to generate a lattice self-energy $\Sigma_{\mathrm{DCA}^+}(\bk, \wn)$ whose coarse-grained average equals the cluster self-energy $\Sigma_c(\bK, \wn)$.
Note that the piecewise constant form of the standard DCA algorithm, Eq.~(\ref{eq:sigma_dca}), trivially satisfies this condition.
Finding the lattice self-energy $\Sigma_\mathrm{DCA^+}(\bk, \wn)$ in \dcaplus is a two step procedure.
First, we interpolate the cluster self-energy $\Sigma_c(\bK, \wn)$ between the cluster momenta $\bK$,
\begin{subequations}
\begin{align}
    \big\{\bK, \Sigma_c(\bK, \wn)\big\} &\rightarrow \big\{ \bk, \tilde{\Sigma}(\bk, \wn) \big\} \,,\\
    \mathrm{with} \quad \tilde{\Sigma}(\bK, \wn) &\equiv \Sigma_c(\bK, \wn) \,.
\end{align}
\end{subequations}
Next, we extend the coarse-graining condition, Eq.~(\ref{eq:coarse-graining_condition_self-energy}), to the interpolated lattice self-energy $\tilde{\Sigma}(\bk, \wn)$,
\begin{equation}
\label{eq:Sigma_deconvolution}
    \tilde{\Sigma}(\bk, \wn) = \frac{N_c}{V_\mathrm{BZ}} \int_\mathrm{BZ} \! d\bk' \, \phi(\bk-\bk') \, \Sigma_{\mathrm{DCA}^+}(\bk', \wn) \,,
\end{equation}
where we used the translation property, Eq.~(\ref{eq:patch_translation_condition}), and inversion symmetry, Eq.~(\ref{eq:patch_inversion_symmetry}), of the patch function $\phi_\bK(\bk)$.
According to Eq.~(\ref{eq:Sigma_deconvolution}), $\tilde{\Sigma}(\bk, \wn)$ is related to $\Sigma_\mathrm{DCA^+}(\bk, \wn)$ by a convolution with the patch function $\phi(\bk-\bk')$.
Consequently, the lattice self-energy $\Sigma_\mathrm{DCA^+}(\bk, \wn)$ is determined by solving the deconvolution problem.
We want to point out that the convergence of both the interpolation and the deconvolution depends on the size of the real-space cluster.
More precisely, if the assumption already the DCA algorithm is based on is violated, i.e. the self-energy of the lattice is longer-ranged, the lattice mapping problem, $\Sigma_c(\bK, \wn) \rightarrow \Sigma_\mathrm{DCA^+}(\bk, \wn)$, is ill-defined and convergence poor~\cite{Staar:2013ec}.
This sensitivity to the behavior of the real-space self-energy defines the current limits of the \dcaplus algorithm.

\begin{figure}
    \centering
    \includegraphics[scale=0.45]{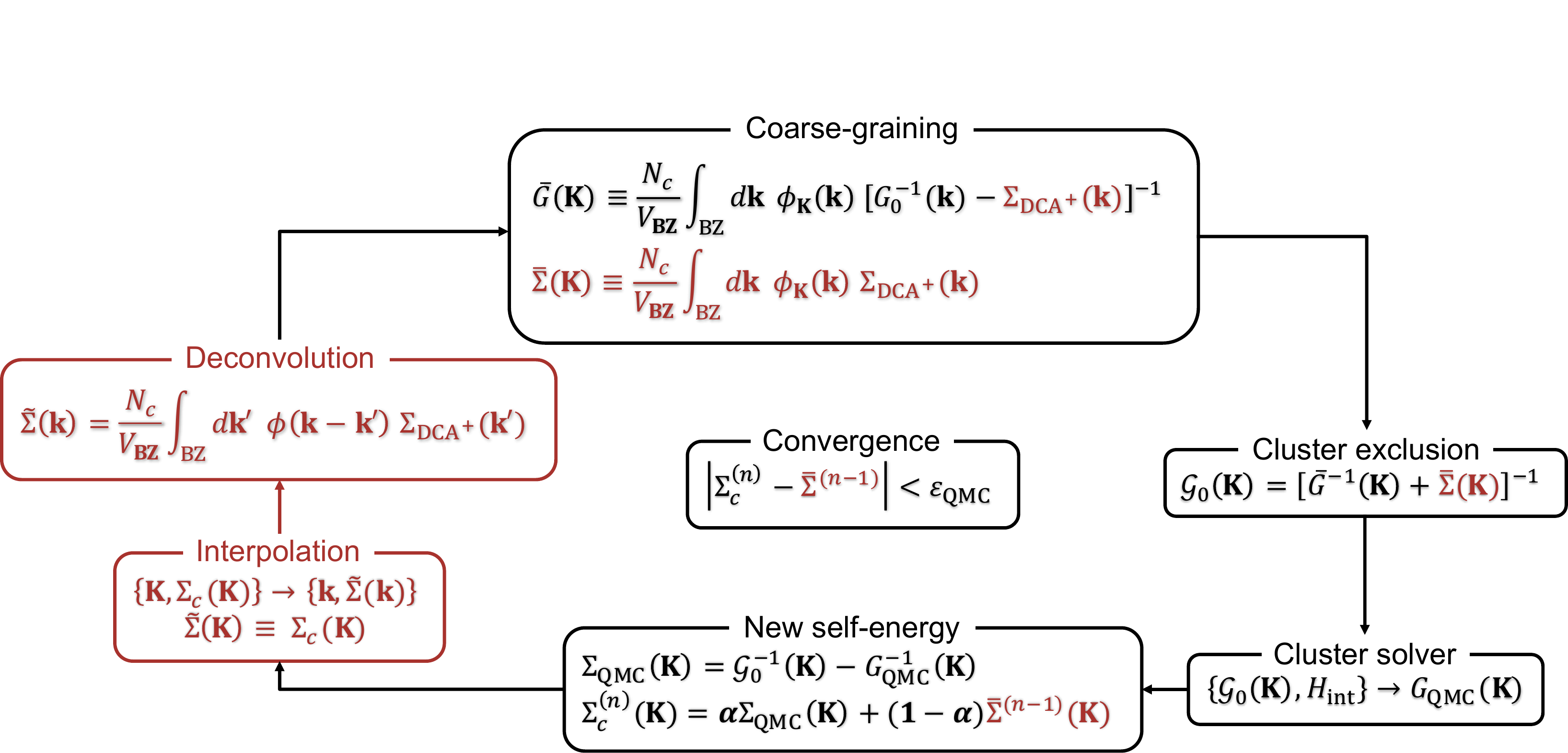}
    \caption{
        (Color online)
        The \dcaplus self-consistency loop with differences to the standard DCA algorithm highlighted in red.
        Due to the non-trivial lattice mapping consisting of an interpolation followed by a deconvolution, the \dcaplus algorithm distinguishes between cluster and coarse-grained self-energy, $\Sigma_c(\bK, \wn)$ and $\bar{\Sigma}(\bK, \wn)$, respectively.
    }
    \label{fig:dcaplus}
\end{figure}

From $\Sigma_{\mathrm{DCA}^+}(\bk, \wn)$ we again calculate the coarse-grained Green's function $\Gb(\bK, \wn)$ and this time also a coarse-grained self-energy $\bar{\Sigma}(\bK, \wn)$, which should be close to the cluster self-energy $\Sigma_c(\bK, \wn)$,
\begin{subequations}
\begin{align}
    \Gb(\bK, \wn) &\equiv \frac{N_c}{V_\mathrm{BZ}} \int_\mathrm{BZ} \! d\bk \, \phi_\bK(\bk) \, \left[ G_0^{-1}(\bk, \wn) - \Sigma_\mathrm{DCA^+}(\bk, \wn) \right]^{-1} \,, \\
    \bar{\Sigma}(\bK, \wn) &\equiv \frac{N_c}{V_\mathrm{BZ}} \int_\mathrm{BZ} \! d\bk \, \phi_\bK(\bk) \, \Sigma_{\mathrm{DCA}^+}(\bk, \wn) \,.
\end{align}
\end{subequations}
The coarse-grained self-energy $\bar{\Sigma}(\bK, \wn)$ enters the calculation of the bare Green's function of the effective cluster problem in \dcaplus,
\begin{equation}
    \mathcal{G}_0(\bK, \wn) = \left[ \Gb^{-1}(\bK, \wn) + \bar{\Sigma}(\bK, \wn) \right]^{-1} \,.
\end{equation}
As in the DCA algorithm, the QMC cluster solver closes the self-consistency loop by generating a new cluster self-energy $\Sigma_c(\bK, \wn)$.
The full \dcaplus algorithm is illustrated in Fig.~\ref{fig:dcaplus}, in which we also point out the differences to the standard DCA loop.

Distinguishing the coarse-grained self-energy $\bar{\Sigma}(\bK, \wn)$ from the cluster self-energy $\Sigma_c(\bK, \wn)$, the \dcaplus algorithm requires a modified convergence criterion.
Self-consistency is reached when the QMC solver produces a cluster self-energy $\Sigma_c^{(n)}$ that agrees, within the statistical Monte Carlo error $\varepsilon_\mathrm{QMC}$, with the coarse-grained self-energy $\bar{\Sigma}^{(n-1)}$ used to set up the cluster problem,
\begin{equation}
    |\Sigma_c^{(n)} - \bar{\Sigma}^{(n-1)}| < \varepsilon_\mathrm{QMC} \,.
\end{equation}

The benefits of the DCA$^+$ algorithm in terms of full momentum resolution and independence of the results on the cluster shape are illustrated in Fig.~\ref{fig:DCA_vs_DCA+_Sigma-band-structure}, where the momentum dependence of the DCA and DCA$^+$ lattice self-energy is shown for the two different 16-site clusters of Fig.~\ref{fig:patches}.
The DCA self-energy is characterized by jump discontinuities between the coarse-graining patches and shows a strong cluster shape dependence in the region between ($\pi, 0$) and ($0, \pi$).
In contrast, the DCA$^+$ algorithm provides a self-energy with smooth momentum dependence and results for the two clusters that agree well.
In addition to curing the cluster shape dependence, the \dcaplus algorithm further reduces the negative sign problem of the underlying QMC method.
This was attributed to the removal of the apparent discontinuities in the DCA self-energy, which had caused artificial long-range correlations \cite{Staar:2013ec}.

\subsection{Continuous-time auxiliary-field quantum Monte Carlo}
\label{subsec:ct-aux}

Developed by Gull {\it et al.}~\cite{Gull:2008cm} specifically for quantum impurity problems, the continuous-time auxiliary field QMC algorithm~(CT-AUX) is the state of the art method to solve the effective cluster problem in DCA and DCA$^+$ calculations.
CT-AUX can treat Hamiltonians containing all types of density-density interactions, including spatially non-local Coulomb repulsion and multiple orbital models, but the algorithm is most useful for large cluster calculations of the single-band Hubbard model.
Instead of describing the algorithm at full length, we want to focus on the relevant parts for a DCA++ user and refer the interested reader to the original paper as well as the detailed review on continuous-time QMC methods given in Ref.~\cite{Gull:2011jd}.

We consider a generalized Hubbard model given by the Hamiltonian
\begin{align}
\label{eq:generalized_hubbard}
    H &= H_0 + H_\mathrm{int} \,, \\
    H_0 &= \sum_{ij, \sigma} t_{ij} c_{i\sigma}^\dagger c_{j \sigma} - \mu \sum_\mu n_\mu + \frac{1}{2} \sum_{\mu > \nu} U_{\mu \nu} \left(n_\mu + n_\nu \right) \,, \\
    H_\mathrm{int} &= \sum_{\mu > \nu} U_{\mu \nu} \left[ n_\mu n_\nu - \frac{1}{2} \left(n_\mu + n_\nu \right) \right] \,.
\end{align}
The indices $i$ and $j$ run over sites and orbitals to allow for the most general type of electron hopping.
$\mu$ and $\nu$ represent combined indices of spin, orbital and site, and the sum $\sum_{\mu > \nu}$ runs over all pairs of correlated spin-orbitals ($U_{\mu \nu} \neq 0$), whose number we denote by $N_\mathrm{corr}$.
Note that for the traditional single-band Hubbard model with only on-site Coulomb repulsion, the number of correlated orbitals is just the number of sites, $N_\mathrm{corr} = N_c$.
Based on this decomposition of the Hamiltonian $H$ into a non-interacting part $H_0$ and the interaction term $H_\mathrm{int}$, we can formulate the partition function $Z = \mathrm{Tr}[ e^{-\beta H}]$ in the corresponding interaction representation,
\begin{equation}
\label{eq:Z_interaction_rep}
    Z = e^{-K} \, \mathrm{Tr} \left[ e^{-\beta H_0} T_\tau e^{-\int_{0}^{\beta}\!d\tau\, \left(H_\mathrm{int}-K/\beta\right)} \right] \,.
\end{equation}
The time-ordered exponential is now expanded with the parameter $K$, introduced in Eq.~(\ref{eq:Z_interaction_rep}), controlling the expansion order,
\begin{equation}
    Z = e^{-K} \sum_{k=0}^\infty \left(\frac{K}{\beta}\right)^k \int_{0}^{\beta}\! d\tau_1 \, \dots \int_{t_{k-1}}^{\beta} \! d\tau_{k} \, \mathrm{Tr} \left[ e^{-\left(\beta-\tau_k+\tau_1\right)H_0} \left(1 - \frac{\beta H_\mathrm{int}(\tau_k)}{K}\right) \dots e^{-\left(\tau_2-\tau_1\right)H_0} \left(1 - \frac{\beta H_\mathrm{int}(\tau_1)}{K}\right) \right] \,.
\end{equation}
The interaction terms $\left(1 - \beta H_\mathrm{int}(\tau)/K\right)$ are decoupled with the help of auxiliary spins $s_i$ \cite{Rombouts:1999ip},
\begin{align}
    1 - \frac{\beta H_\mathrm{int}}{K} &= \frac{1}{2 N_\mathrm{corr}} \sum_{\mu > \nu} \sum_{s=\pm 1} e^{\gamma_{\mu \nu} s \left(n_\mu - n_\nu \right)} \,, \\
    \cosh(\gamma_{\mu \nu}) &= 1 + \frac{U_{\mu \nu} \beta N_\mathrm{corr}}{2 K} \,.
\end{align}
As a result, the partition function $Z$ becomes a high-dimensional integral,
\begin{align}
    Z       &= e^{-K} \sum_{k=0}^\infty \sum_{\mu > \nu} \sum_{s_1, \dots, s_k=\pm 1} \int_{0}^{\beta} \! d\tau_1 \, \dots \int_{\tau_{k-1}}^{\beta} \! d\tau_k \left( \frac{K}{2 \beta N_\mathrm{corr}} \right)^k Z_k(\left\{\mu, \nu, \tau, s\right\}_k) \,, \\
    Z_k(\left\{\mu, \nu, \tau, s\right\}_k) &=\mathrm{Tr} \prod_{i=1}^{k} e^{-\Delta \tau_i H_0} e^{\gamma_{\mu \nu} s (n_\mu - n_\nu)} = Z_0 \prod_{\sigma} \det N_\sigma^{-1}(\left\{\mu, \nu, \tau, s, \right\}_k) \,, \label{eq:Zk}
\end{align}
which can be evaluated by a Markov Chain Monte Carlo procedure.

\begin{figure}[h]
    \centering
    \includegraphics[scale=0.45]{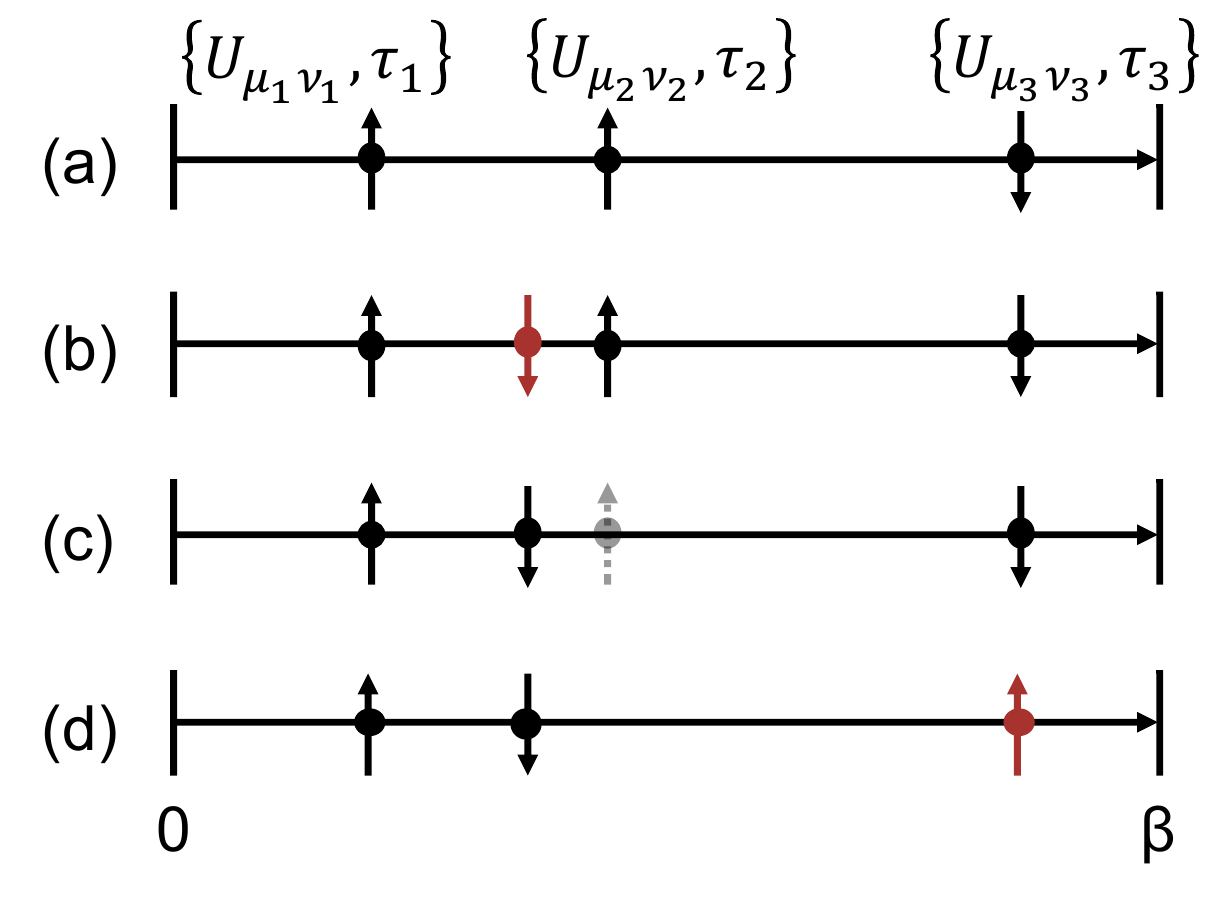}
    \caption{
        (Color online)
        CT-AUX vertex configuration of a $k=3$ order term (a) and the three types of single spin updates:
        (a) to (b) vertex insertion, (b) to (c) vertex removal, (c) to (d) flip of an auxiliary spin.
    }
    \label{fig:ct-aux}
\end{figure}

Following the Metropolis algorithm, we sample the configuration space $\mathcal{C}$, in which an element $c \in \mathcal{C}$ consists of a set of $k$ vertices, $\{v = (\mu_i, \mu_j, \tau, s)\}_k$.
Each vertex possesses an auxiliary spin $s = \pm 1$ and represents an interaction between two correlated spin-orbitals, $\mu_i = \left\{\sigma_i, \alpha_i, \br_i\right\}$ and $\mu_j = \left\{\sigma_j, \alpha_j, \br_j\right\}$, acting at a time $\tau \in [0, \beta)$.
A configuration $c \in \mathcal{C}$ in the Markov chain is updated by inserting a new vertex, removing an existing vertex, or flipping an auxiliary spin.
These single spin updates correspond to rank-1 operations on the $N_\sigma$-matrices defined in Eq.~(\ref{eq:Zk}).
The dimension of the $N_\uparrow(N_\downarrow)$-matrix is given by the number of $\uparrow (\downarrow)$-electron spins contained in the current configuration.
If interactions only occur between spins of opposite type, both $N_\uparrow$ and $N_\downarrow$ are $k \times k$ matrices.
Fig.~\ref{fig:ct-aux} shows a typical configuration of auxiliary spins on the imaginary time axis and the three types of single spin updates.

While we sample configurations on the imaginary time axis, measuring the Green's function is most accurately performed in discrete Matsubara frequency space.
Time measurements would lead to binning errors as the vertices can occupy arbitrary positions in the imaginary time interval.
From a vertex configuration $\{v\}_k$ and its respective $N_\sigma$-matrices we first compute a function $M$, whose Fourier coefficients are then accumulated.
The Green's function is linearly related to $M$,
\begin{equation}
    \langle G(\bk, \wn) \rangle = \mathcal{G}_0(\bk, \wn) - \frac{1}{\beta} \mathcal{G}_0(\bk, \wn) \, \langle M(\bk, \wn) \rangle \, \mathcal{G}_0(\bk, \wn) \,.
\end{equation}
The self-energy $\Sigma(\bk, \wn)$ is finally gained from the measured Green's function $\langle G(\bk, \wn) \rangle$ and the bare propagator of the cluster impurity problem $\mathcal{G}_0(\bk, \wn)$ using the Dyson equation,
\begin{equation}
\label{eq:dyson_sigma}
    \Sigma(\bk, \wn) = \mathcal{G}_0^{-1}(\bk, \wn) - \langle G(\bk, \wn) \rangle ^{-1} \,.
\end{equation}
In addition to the single-particle Green's function $G(\bk, \wn)$, also the two-particle Green's function can be obtained within the CT-AUX algorithm.
The measurement of the latter involves significantly higher computational cost, though.

The CT-AUX algorithm scales as the cube of the linear size of the $N_\sigma$-matrices, i.e. $\mathcal{O}(\langle k \rangle^3)$.
The average expansion order $\langle k \rangle$, in turn, grows linearly with the expansion parameter $K$, the inverse temperature $\beta$, the strength the interactions $U_{\mu \nu}$, and the number of interaction terms $N_\mathrm{corr}$.
In simulations of the Hubbard model away from half-filling, CT-AUX exhibits a negative sign problem.
The signal-to-noise ratio decreases exponentially when the system size or the strength of the Coulomb interaction is increased, or the temperature lowered.
Therefore, the behavior of both the average expansion order and the sign problem makes CT-AUX calculations of large systems, with strong coupling and at low temperatures computationally most demanding.

\paragraph*{Submatrix updates}
An important improvement in terms of computational efficiency of the CT-AUX algorithm has been achieved with the introduction of \emph{submatrix updates}~\cite{Gull:2011hh}.
Combining $k_s$ successive single spin updates into a rank-$k_s$ operation, submatrix updates substantially improve data locality and thus benefit from the cache hierarchy of modern processors.

The formalism of submatrix updates assumes that individual spins are changed only once within a submatrix update and spins that do not satisfy this condition require special treatment at the cost of a small performance loss.
One can either update these spins with the help of Bennett's algorithm~\cite{Bennett:1965dd} or abort the current submatrix update when a spin is selected twice and propose the same spin again at the beginning of the next update.
For systems with large expansion orders, for which the probability of touching the same spin twice within a submatrix update is small, ignoring proposals of already chosen spins is a valid option, too.
This approximation avoids the performance loss, but violates the detailed balance condition.

\paragraph*{Non-equidistant FFT measurements}
A remarkable speed-up in the measurement of single- as well as the two-particle functions has been accomplished by developing optimized non-equispaced fast Fourier transform~(NFFT) methods~\cite{Staar:2012fj}.
In the context of CT-AUX, the methods are used to measure and accumulate the $M$ function, but their applicability extends to other continuous-time QMC methods as well.

Since vertex configurations are not sampled on an equispaced grid but at random time locations, we cannot directly apply the fast Fourier transform~(FFT) algorithm on the raw samples.
To circumvent this problem, the NFFT algorithm~\cite{Keiner:2009js} makes use of the convolution theorem,
\begin{equation}
\label{eq:convolution_theorem}
    f_\omega \, \varphi_\omega = \int_0^\beta \! d\tau \, e^{i \omega \tau} \left[ \int_{0}^{\beta} \! d\tau' \, \varphi(\tau-\tau') f(\tau') \right] \,.
\end{equation}
Given samples $f_i$ at random times $\tau_i \in [0, \beta)$, we first transform the data-set
onto an equispaced imaginary time grid by convoluting it with a localized, $\beta$-periodic kernel $\varphi$.
We are then able to use the FFT algorithm to compute the Fourier transform of the convoluted data.
According to Eq.~(\ref{eq:convolution_theorem}), we obtain the Fourier components $f_\omega$ of the original data after normalizing with the Fourier coefficients of the kernel $\varphi_\omega$, which finalizes the NFFT algorithm,
\begin{equation}
\label{eq:nfft}
    f_\omega \leftarrow \frac{1}{\varphi_\omega} \text{FFT} \left[ \left\{ \bar{f_l} = \sum_i \varphi(\tau_i - \frac{l \beta}{m N_\omega}) f_i \, | \, l \in 0, 1, \dots, m N_\omega-1 \right\} \right] \,.
\end{equation}
In Eq.~(\ref{eq:nfft}), $N_\omega$ denotes the desired number of positive Matsubara frequencies and $m$ is an oversampling factor.

The linearity of the Fourier transform allows us to delay it, meaning we can accumulate the equispaced data points $\bar{f}_l$ and perform a single FFT on the accumulated data at the end of the Monte Carlo integration.
The algorithm can be further accelerated by approximating the kernel $\varphi$ with a low order polynomial at the cost of a controllable interpolation error.

While the delayed-NFFT scheme cannot be applied to the measurement of the two-particle Green's function as Fourier transformation and accumulation are not interchangeable there, a specifically tuned 2D NFFT algorithm can still lead to a considerable speed-up compared to the standard 2D non-equidistant discrete Fourier transform~\cite{Staar:2012fj}.

\subsection{Continuous-time hybridization expansion quantum Monte Carlo}
\label{subsec:ct-hyb}

With the continuous-time hybridization expansion QMC algorithm~(CT-HYB)~\cite{Werner:2006iz, Werner:2006ko}, DCA++ provides an alternative continuous-time impurity solver.
While CT-AUX is the favorable algorithm for large cluster DCA and \dcaplus simulations, CT-HYB is the method of choice for single-site multi-orbital impurity models in the strong coupling regime ($U/t \approx 10-\infty$).
The CT-HYB algorithm is capable of treating general interactions, but provides a particularly efficient \emph{segment formulation} for density-density interactions.
We will restrict ourselves to this formulation and again refer readers to the original papers and Ref.~\cite{Gull:2011jd} for a complete discussion.

\begin{figure}[h]
    \centering
    \begin{minipage}[t]{0.48\columnwidth}
        \centering
        \includegraphics[scale=0.45]{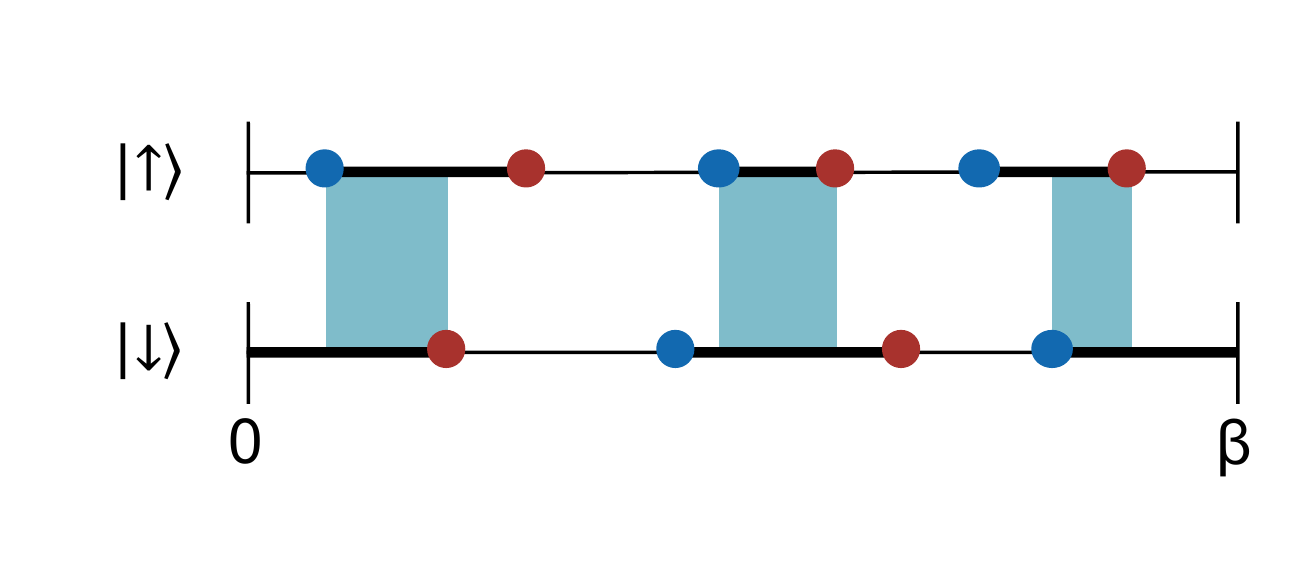}
        \caption{
            (Color online)
            A segment-CT-HYB configuration of a $k=5$ order term for a single-orbital impurity.
            Thick lines represents segments, which are bounded by creation (blue dots) and annihilation operators (red dots).
            As shown for the spin-down orbital, segments can wrap around $\beta$.
            Shaded areas indicate time intervals in which both spin-orbitals are occupied and the Coulomb interaction $U$ is acting between them.
        }
        \label{fig:ct-hyb-configuration}
    \end{minipage}
    \hspace{0.03\columnwidth}
    \begin{minipage}[t]{0.48\columnwidth}
        \centering
        \includegraphics[scale=0.45]{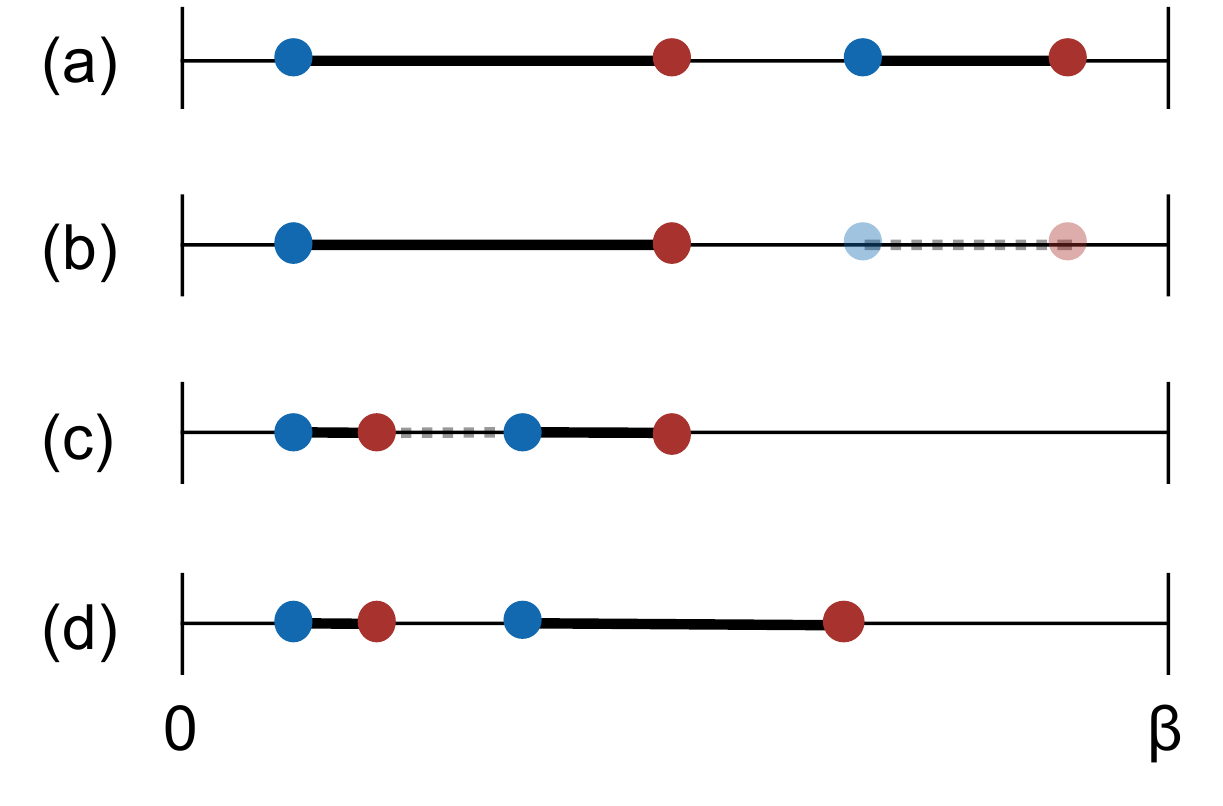}
        \caption{
            (Color online)
            Three examples of segment updates:
            (a) to (b) segment removal, (b) to (c) anti-segment insertion, (c) to (d) shift of segment endpoint.
        }
        \label{fig:ct-hyb-segment-updates}
    \end{minipage}
\end{figure}

We consider a quantum impurity model, which describes a small system, the impurity, embedded in an infinite non-interacting bath,
\begin{equation}
    H = H_\mathrm{imp} + H_\mathrm{bath} + H_\mathrm{hyb} \,.
\end{equation}
According to its name, the CT-HYB algorithm is based on an expansion of the partition function $Z$ in the hybridization $H_\mathrm{hyb}$, which couples the impurity to the bath.
In analogy to the procedure in CT-AUX, the partition function is transformed into a high-dimensional integral, which is sampled by means of a Markov chain Monte Carlo simulation.
In the segment picture the configuration space $\mathcal{C}$ is represented by multiple imaginary time lines.
Each line spans the interval $[0, \beta)$ and corresponds to a flavor (spin, orbital, site) of the impurity.
A configuration $c \in \mathcal{C}$ consists of a set of segments on these time lines during which the corresponding spin-orbitals are occupied.
For the Markov chain to be ergodic, two types of updates are needed: segment insertions and removals.
Additional updates in form of insertion and removal of ``anti-segments'' and segment shifts, however, can improve the sampling efficiency.
Segment swaps, which exchange the time lines of two flavors, may be necessary in certain cases.
A typical configuration and examples of  segment updates are illustrated in Figs.~\ref{fig:ct-hyb-configuration} and \ref{fig:ct-hyb-segment-updates}, respectively.

The CT-HYB algorithm accumulates the Matsubara coefficients of the Green's function $G(\wn)$.
In addition, it has proven useful to measure $G(\wn) \Sigma(\wn)$.
Although it is not necessary to accumulate this product, since the self-energy can be obtained from the Dyson equation~(\ref{eq:dyson_sigma}), the approach leads to much smaller statistical errors~\cite{Hafermann:2012cg}.

For a diagonal hybridization term $H_\mathrm{hyb}$, the CT-HYB algorithm in the segment formulation is sign problem free and scales as $\mathcal{O}(N \beta^3)$, where $N$ is the number of spin-degenerate orbitals.

\subsection{Calculation of two-particle correlation functions}
\label{subsec:susceptibilities}

A possible approach to study phase transitions, for example to the superconducting state, is the extension of the DCA and DCA$^{+}$ frameworks to the two-particle level.
In this section we want to briefly outline this formalism for the case of the particle-particle channel and refer to the article by Jarrel {\it et al.}~\cite{Jarrell:2001jj} for a more detailed derivation in the DCA framework, to the work of Staar, Maier, and Schulthess~\cite{Staar:2014gz} for the extension to DCA$^+$, and to the didactic summary by Maier~\cite{Maier:2015ju}, which we follow here.

\paragraph*{DCA}
We start from the Bethe-Salpeter equation~(BSE), which in the particle-particle channel (abbreviated with ``pp'') is given by
\begin{equation}
\label{eq:bse_lattice}
    G_2(k, q-k, q-k', k') = G_\uparrow(k) G_\downarrow(q-k) \delta_{k, k'} - \frac{T}{N} \sum_{k''} G_\uparrow(k) G_\downarrow(q-k) \, \Gamma^{pp}(k, q-k, q-k'', k'') \, G_2(k'', q-k'', q-k', k') \,.
\end{equation}
It relates the two-particle Green's function $G_2$ to the irreducible vertex function of this channel $\Gamma^{pp}$, and can be regarded as the two-particle level analog of the Dyson equation.
In Eq.~(\ref{eq:bse_lattice}) we introduced a combined index for momentum and Matsubara frequency, $k \equiv (\bk, \wn)$, with the corresponding summation, $\frac{T}{N}\sum_k \equiv \frac{T}{V_\mathrm{BZ}} \int d\bk \sum_\wn$. The variable $q \equiv (\bq, \vm)$ represents the transferred momentum and transferred (bosonic) frequency.
Applying the underlying assumption of the DCA at the two-particle level, we expand the irreducible vertex function of the cluster onto the coarse-graining patches to obtain a piecewise constant approximation of the lattice irreducible vertex function,
\begin{equation}
\label{eq:Gamma_dca}
    \Gamma_\mathrm{DCA}^{pp}(k, k', q) \equiv \sum_{\bK, \bK'} \phi_\bK(\bk) \, \Gamma_c^{pp}(K, K', q) \, \phi_{\bK'}(\bk') \,,
\end{equation}
where $\Gamma^{pp}(k, k', q) \equiv \Gamma^{pp}(k, q-k, q-k', k')$ was used to simplify the notation.
The cluster vertex function $\Gamma_c^{pp}(K, K', q)$, in turn, satisfies the cluster version of the Bethe-Salpeter equation~(\ref{eq:bse_lattice}),
\begin{equation}
\label{eq:bse_cluster}
    G_{2, c}(K, K', q) = G^{0,pp}_{2, c}(K, K', q) - \frac{T}{N_c} \sum_{K''} G^{0,pp}_{2, c}(K, K', q) \, \Gamma_c^{pp}(K, K'', q) \, G_{2, c}(K'', K', q) \,,
\end{equation}
with the non-interacting two-particle Green's function $G^{0, pp}_{2, c}(K, K', q) \equiv G_{c, \uparrow}(K) G_{c, \downarrow}(q-K) \delta_{K, K'}$.
Solving for the irreducible vertex function yields (matrix notation in $K, K'$ for fixed $q$),
\begin{equation}
    \mathbf{\Gamma}_c^{pp} = -\frac{N_c}{T} \left[ \left(\mathbf{G}^0_{2, c}\right)^{-1} - \left(\mathbf{G}_{2, c}\right)^{-1} \right] \,.
\end{equation}
The single- and two-particle Green's functions are obtained from the cluster solver.

Possible phase transitions can be discovered from the eigenvalues and left and right eigenvectors of the Bethe-Salpeter kernel,
\begin{align}
    -\frac{T}{N} \sum_{k} \tilde{g}_\alpha(k) \, \Gamma^{pp}(k, k', q) G_\uparrow(k') G_\downarrow(q-k') &= \lambda_\alpha \tilde{g}_\alpha(k') \,, \\
    -\frac{T}{N} \sum_{k'} \Gamma^{pp}(k, k', q) G_\uparrow(k') G_\downarrow(q-k') \, g_\alpha(k') &= \lambda_\alpha g_\alpha(k) \,, \label{eq:bse_eig_lattice}    
\end{align}
where we dropped the dependence of $\lambda_\alpha$, $g_\alpha(k)$ and $\tilde{g}_\alpha(k)$ on the transferred momentum and frequency $q$, as well as on the considered channel.
This can be seen when the two-particle Green's function is expressed in terms of these eigenvalues and eigenvectors,
\begin{equation}
\label{eq:G2_eig}
    G_2^{pp}(k, k', q) = \sum_{\alpha} G_\uparrow(k) G_\downarrow(q-k) \frac{g_\alpha(k) \tilde{g}_\alpha(k')}{1-\lambda_\alpha} \,.
\end{equation}
The eigenvalues $\lambda_\alpha$ indicate an instability of the system: Eq.~(\ref{eq:G2_eig}) diverges when the leading eigenvalue approaches 1. The eigenvectors $g_\alpha(k)$ and $\tilde{g}_\alpha(k)$ describe the momentum and frequency structure of the instability.

By using the approximation of the lattice irreducible vertex function by step functions, Eq.~(\ref{eq:Gamma_dca}), and restricting the momentum resolution of the right eigenvectors $g_\alpha(\bk, \wn)$ to the cluster momenta $\bK$, we can transform Eq.~(\ref{eq:bse_eig_lattice}) into a coarse-grained eigenvalue equation,
\begin{equation}
\label{eq:BS_kernel_cluster}
    -\frac{T}{N_c} \sum_{K'} \Gamma_c^{pp}(K, K', q) \, \chi_0^{pp}(K', q) \, g_\alpha(K') = \lambda_\alpha g_\alpha(K) \,,
\end{equation}
with
\begin{equation}
    \chi^{pp}_0(K, q) \equiv \frac{N_c}{V_\mathrm{BZ}} \int \! d\bk \, \phi_\bK(\bk) \, G_\uparrow(k) G_\downarrow(q-k) \,.
\end{equation}
Eq.~(\ref{eq:BS_kernel_cluster}) is easily solvable by standard diagonalization techniques.

\paragraph*{\dcaplus} To restore the momentum resolution of the lattice, we need to find a lattice irreducible vertex function $\Gamma^{pp}(k, k', q)$ that is continuous in $\bk$ and $\bk'$.
In analogy to the derivation of \dcaplus in Section~\ref{subsec:dca_plus}, we start from a coarse-graining equation for the vertex function,
\begin{equation}
    \Gamma^{pp}_c(K, K', q) = \frac{N_c^2}{V_\mathrm{BZ}^2} \int \! d\bk d\bk' \, \phi_\bK(\bk) \, \Gamma^{pp}(k, k', q) \, \phi_{\bK'}(\bk') \,,
\end{equation}
which requires the coarse-grained average of the lattice function to match the corresponding cluster quantity.
Consistent with the computation of the continuous lattice self-energy $\Sigma_{\mathrm{DCA}^+}(\bk,
\wn)$, we obtain $\Gamma^{pp}_{\mathrm{DCA}^+}(k, k', q)$ from an interpolation of the cluster vertex function,
\begin{equation}
    \left\{ \bK, \Gamma^{pp}_c(K, K', q) \right\} \rightarrow \left\{ \bk, \tilde{\Gamma}^{pp}(k, k', q) \right\} \,,
\end{equation}
followed by a deconvolution,
\begin{equation}
\label{eq:gamma_deconvolution}
    \tilde{\Gamma}^{pp}(k_1, k_2, q) = \frac{N_c^2}{V_\mathrm{BZ}^2} \int \! d\bk_1' d\bk_2' \, \phi(\bk_1-\bk_1') \, \Gamma^{pp}_{\mathrm{DCA}^+}(k_1', k_2', q) \, \phi(\bk_2-\bk_2') \,.
\end{equation}
From $\Gamma_{\mathrm{DCA}^+}^{pp}(k, k', q)$ we can compute the Bethe-Salpeter eigenvectors and eigenvalues with full momentum resolution (see Eq.~(\ref{eq:bse_eig_lattice})).

In some cases it has proven useful to project the Bethe-Salpeter kernel onto a set of crystal harmonics corresponding to localized lattice vectors.
This \emph{folding} not only leads to much smaller matrices to diagonalize, but more importantly reduces noise by projecting onto the physically relevant subspace.

\section{The DCA++ code}
\label{sec:code}

\subsection{Main components}
\label{subsec:overview}

\begin{figure}[h]
    \centering
    \includegraphics[width=\columnwidth]{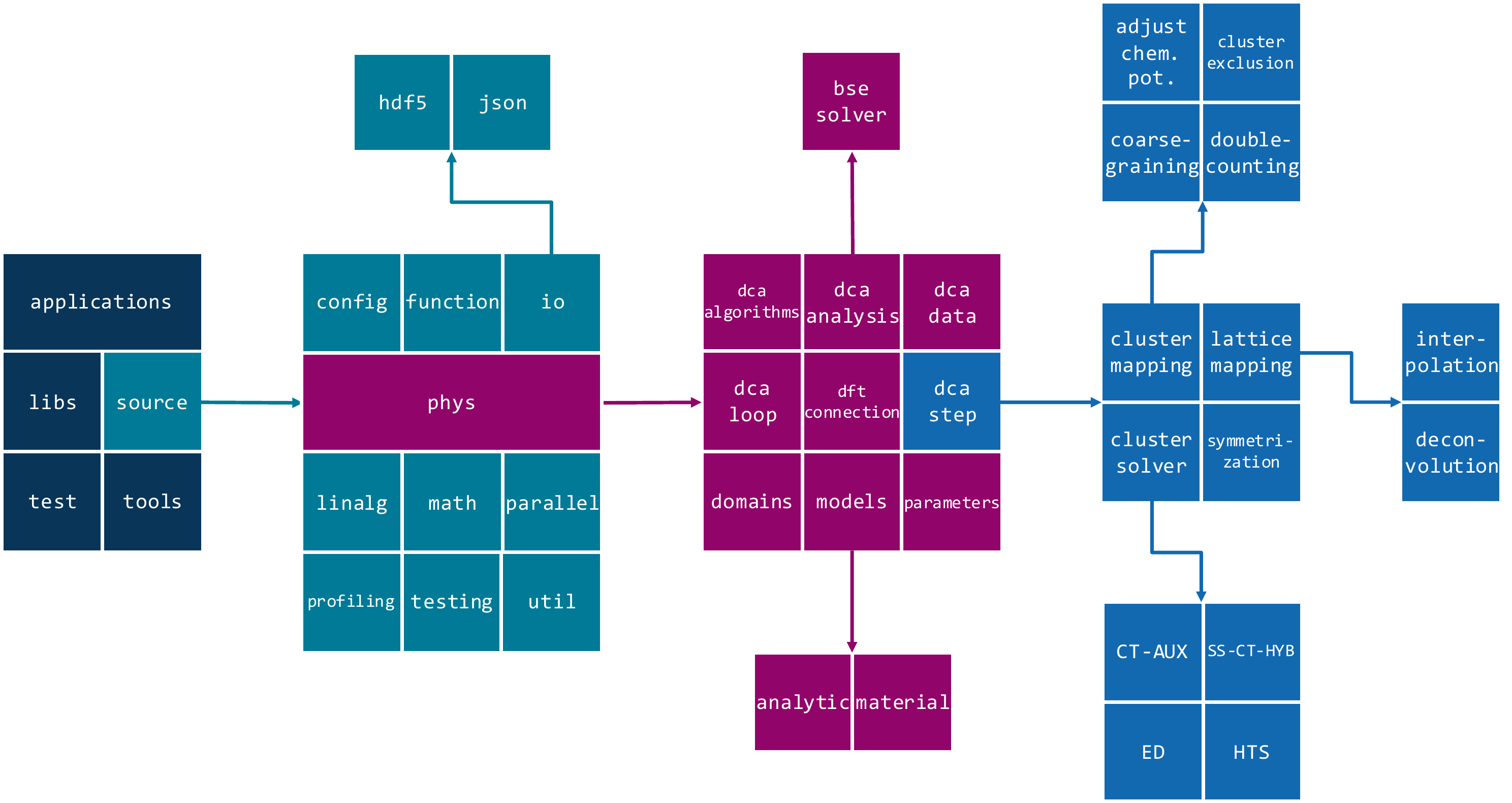}
    \caption{
        (Color online)
        Schematic overview of the code base.
        The modular and hierarchical structure reflects the generic design of the DCA++ code.
    }
    \label{fig:dca++_overview}
\end{figure}

The main purpose of this section is to give an overview of the key parts and main capabilities of the DCA++ code.
A visual summary of the code base is shown in Fig.~\ref{fig:dca++_overview}.

We have chosen C++ as the primary implementation language for its generic programming model, which, besides easy extensibility, allows us to hide architectural details.
The C++14 standard, which DCA++ adopts, provides many modern language features for effective and safe programming, while still having a wide compiler support.
On top of C++, we use CUDA to utilize GPU resources in performance critical parts of the code.

\paragraph*{Cluster mapping}
Distinguishing DCA and DCA$^+$ from finite-size methods, the coarse-graining procedure is a central element of the code.
In addition to the traditional coarse-graining, which partitions the first Brillouin zone according to the Brillouin zones of the superlattice, the interlaced coarse-graining developed by Staar {\it et~al.}~\cite{Staar:2016ia} is implemented with adjustable number of periods, that is variable degree of interleaving.
Besides the coarse-graining routines, the cluster mapping module contains methods to update the chemical potential and execute the cluster-exclusion step.
In LDA+DMFT calculations the double-counting problem is addressed by the correspondent submodule with a constant correction term~\cite{Kotliar:2006fl}, specified as an input parameter.

\paragraph*{Lattice mapping} The framework of \dcaplus requires a non-trivial lattice mapping to generate a lattice self-energy and irreducible vertex function with full momentum resolution.
After interpolation of the cluster functions onto the fine lattice grid, the lattice mapping becomes a deconvolution problem.
In the case of the self-energy, we employ the iterative Richardson-Lucy deconvolution algorithm as explained in \cite{Staar:2013ec}.
For the deconvolution of the irreducible vertex function we have chosen a simpler approach. After Fourier transformation of the convolution equation~(\ref{eq:gamma_deconvolution}) to real space, the deconvolution problem presents itself as a simple division.

\paragraph*{Cluster solver}
Solving the effective cluster problem and computing a new estimate of the cluster self-energy is the most time consuming part in a DCA or \dcaplus calculation.
For this reason, lots of effort has been put in the development of a cutting edge implementation of the CT-AUX algorithm including submatrix updates and efficient NFFT measurements.
With the aim of featuring optimal methods for the entire spectrum of quantum impurity models, we provide a complementary cluster solver to CT-AUX in the form of the segment formulation of CT-HYB restricted to single-site problems~(SS-CT-HYB).
Besides the two continuous-time QMC methods, DCA++ implements two other cluster solvers, each bound to a special purpose.
A high temperature series expansion~(HTS) solver may help improve convergence of the lattice mapping by computing a fourth order perturbation expansion of the self-energy with respect to the on-site Coulomb interaction $U$.
To validate QMC methods and new models, an ED solver can be employed to compute exact results on small lattices in finite-size calculations (no mean-field).

\paragraph*{Models}
DCA++ is able to simulate standard, extended and frustrated Hubbard models on various types of lattices.
2D square and triangular lattices, and a bilayer square lattice corresponding to a two-orbital model are currently supported, but new lattices and models can easily be added upon need.
In particular, the code's generic structure based on C++ templates permits to easily treat 1D and 3D lattices, too.
In the framework of LDA+DMFT real materials such as CuO$_2$ and NiO can be studied in ab initio calculations.

\paragraph*{Input/output}
Running simulations requires reading input files and writing out results.
The input and output~(I/O) module supports two data formats, HDF5~\cite{hdf5} and JSON~\cite{json}.
While JSON's human-readability makes it very suitable for specifying input parameters, we recommend using the binary HDF5 format to write output.
HDF5 features high speed in reading and writing data, and requires less storage size compared to other data formats.

\begin{figure}[h]
    \centering
    \includegraphics[scale=0.45]{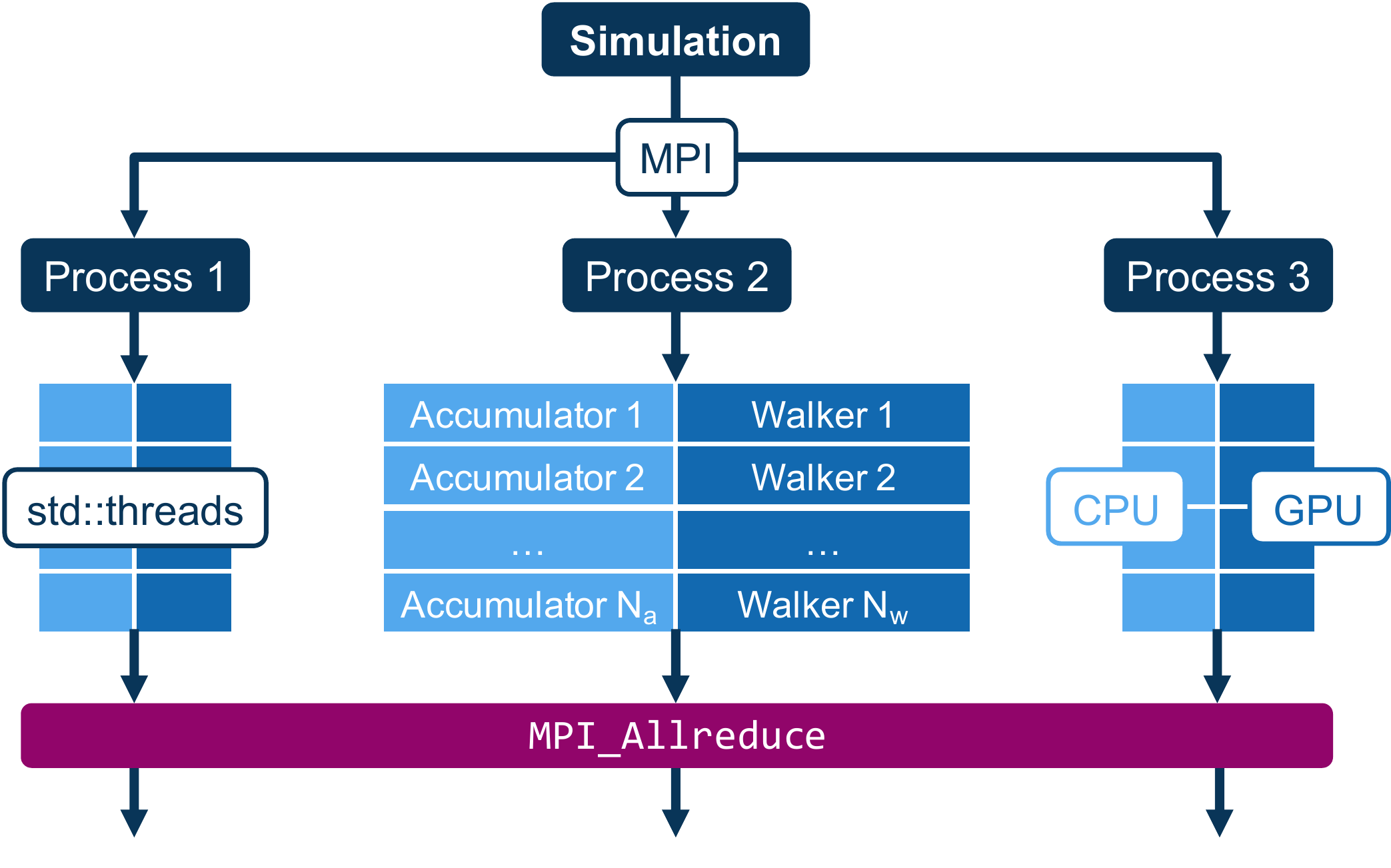}
    \caption{
        (Color online)
        MPI + threading parallelization scheme of a DCA or \dcaplus calculation.
        If GPU resources are available, the computationally heavy matrix operations performed by a CT-AUX walker are executed on them.
        Communication between MPI processes is limited to a few \texttt{MPI\_Allreduce} calls at the very end of the Monte Carlo integration.
    }
    \label{fig:parallelization}
\end{figure}

\paragraph*{Parallelization}
The computationally most expensive part of a DCA or \dcaplus calculation is the QMC step.
Fortunately, Monte Carlo simulations are embarrassingly parallel, which we exploit on distributed multi-core machines with a two level (MPI + threading) parallelization scheme, illustrated in Fig.~\ref{fig:parallelization}.
Building DCA++ with MPI enabled allows to run the DCA or \dcaplus calculation with multiple MPI processes.
Each MPI process performs an independent Monte Carlo integration and results are accumulated at the end in few collective reductions.
In addition, a threaded Monte Carlo wrapper allows each MPI process to spawn multiple Monte Carlo walker and accumulator threads.
Each Monte Carlo walker processes its own Markov chain and, as soon as it has executed the required number of updates per measurement, sends its current configuration over to a queue of idle Monte Carlo accumulators.
A free accumulator then takes measurements on the sampled configuration.
The total number of measurements during the Monte Carlo step of a single DCA or \dcaplus iteration is given by
\begin{align*}
    \text{total number of measurements} =
        &\text{ number of MPI processes} \\
        &\times \text{ number of accumulators} \\
        &\times \text{ number of measurements per accumulator}\,.
\end{align*}

\paragraph*{GPU}
In addition to conventional multi-core architectures, DCA++ ports to the emerging hybrid CPU-GPU systems.
Comprising large matrix-matrix multiplications, the CT-AUX cluster solver with submatrix updates particularly benefits from the computing power of the GPU.
If DCA++ is built with GPU support enabled, all Monte Carlo walkers of a process are moved to its dedicated GPU.
This leads to a considerable speed-up compared to the multi-core implementation~\cite{Staar:2013ik}.

\paragraph*{Applications}
\texttt{main\_dca} is the central application of DCA++.
An input parameter allows to choose at runtime whether the DCA or \dcaplus algorithm should be executed.
One can turn off the mean-field and do a finite-size QMC calculation, too.
Depending on input parameters, either a fixed number of iterations are carried out, or the DCA or \dcaplus algorithm is iterated until the desired accuracy is reached.
The main output of the application is the cluster's single-particle Green's function and self-energy.
The lattice self-energy can be written out, too, when \dcaplus is used.
If required, the application additionally measures and stores the two-particle Green's function of the cluster.

The output of \texttt{main\_dca} is post processed by the application \texttt{main\_analysis}.
This application computes the eigenvectors and eigenvalues of the Bethe-Salpeter equation, which provide information about nature and extent of instabilities in the system and possible phase transitions.
Like \texttt{main\_dca}, \texttt{main\_analysis} can be run in DCA or \dcaplus mode.
While in DCA mode the momentum resolution of the BSE eigenvectors is restricted to the cluster momenta, the BSE solver produces eigenvectors with continuous momentum dependence when \dcaplus is used.

\subsection{Building}

\paragraph*{Prerequisites}
Building DCA++ requires a set of tools and libraries:
\begin{itemize}
\item a C++14 compiler (tested: clang $\geq$ 3.5, gcc $\geq$ 4.9),
\item the CMake build system~\cite{Martin:2027837} version 3.3 or higher,
\item the HDF5 library with the C++ interface (tested: 1.8.13 and 1.10),
\item FFTW3 \emph{or} an FFT library with the FFTW3 interface (e.g. Intel MKL),
\item BLAS and LAPACK libraries, and
\item MPI, if requested.
\end{itemize}

To enable the GPU support, one needs to provide
\begin{itemize}
\item the CUDA Toolkit (tested: 7.0, 7.5 and 8.0), and
\item the MAGMA library (tested: 2.0.0 with CUDA 7.0 and 2.2.0 with CUDA 7.5 and 8.0).
\end{itemize}

Further optional requirements are
\begin{itemize}
\item the EasyBuild framework~\cite{Hoste:2013hh},
\item Python, NumPy, SciPy, Matplotlib, and the h5py package to use the provided python scripts.
\end{itemize}

\paragraph*{Building steps}
The DCA++ project is maintained in a public GitHub repository at \url{https://github.com/CompFUSE/DCA}.
Cloning the repository allows to obtain the latest version of the master branch.
Alternatively, one can download a specific version from the \href{https://github.com/CompFUSE/DCA/releases}{release page}.
The recommended way of configuring and building DCA++ across all platforms is to use the CMake build system.
The building procedure starts with creating a clean build directory.
The \texttt{cmake} command followed by the path to the source and appropriate options generates the build files.
Finally, by evoking \texttt{make} one can compile the applications and tests, if enabled.
The building steps are summarized in Listing~\ref{listing:build}.
More details on the building procedure can be found on the correspondent \href{https://github.com/CompFUSE/DCA/wiki/Building}{Wiki page} in the GitHub repository, while a complete list of all relevant CMake options with descriptions and their default value is provided in \ref{app:cmake_options}.

\begin{lstlisting}[
caption={Summary of the DCA++ building steps. The \texttt{cmake} command requires the path to the source (here \texttt{../dca\_source}) and can be complemented by a list of options using the \texttt{-D} flag. If tests have been built, they can be executed with the command \texttt{make test}.},
captionpos=b,
label=listing:build]
$ git clone <@{https://github.com/CompFUSE/DCA.git}@> dca_source
$ mkdir build && cd build
$ cmake ../dca_source -D<variable1>=<value1> -D<variable2>=<value2> ...
$ make
$ make test
\end{lstlisting}

\subsection{Running}
\label{subsec:running}

\paragraph*{Input files}
There are two types of input files.
All applications read \emph{simulation parameters}, e.g. the DCA cluster, the temperature or output filenames, from a JSON-formatted input file, in which parameters are thematically grouped and sub-grouped in JSON objects.
\ref{app:parameters} provides descriptions of all input parameters including their type and default value.

In addition to providing simulation parameters, \emph{data} needs to be transferred between individual runs.
A \texttt{main\_dca} run uses the output of a previous calculation to initialize the self-energy with a non-zero value (see below) and the application \texttt{main\_analysis} requires a \texttt{main\_dca} output file containing the cluster single-particle and two-particle Green's functions, as well as the self-energy.
While the file format for data can either be HDF5 (recommended) or JSON, it has to match between consecutive runs.

\paragraph*{Parallelization}
On modern systems with multi-core nodes we usually run up to two MPI tasks per node and one Monte Carlo walker or accumulator thread per core.
The optimal distribution of Monte Carlo walker and accumulator threads, on the other hand, is strongly problem dependent.
Note that memory restrictions on the GPU might limit the number of walker threads a process can spawn.
The environment variable \texttt{OMP\_NUM\_THREADS} can be used to enable threading in BLAS, LAPACK and FFTW routines.
One should be aware, however, that some of the calls to these routines are already within parallel regions.

\paragraph*{Cooldown}
A full DCA or \dcaplus calculation is usually performed by slowly \emph{cooling down} the system.
In a series of \texttt{main\_dca} runs the temperature is gradually lowered to study its dependence on fluctuations leading to phase transitions.
Sometimes this procedure is also necessary to guarantee convergence of the DCA and DCA$^+$ loop at low temperatures.

The calculation is started at a high temperature ($T \approx 1.0 t$), at which a vanishing self-energy is a good initial guess, and the system then slowly cooled down.
Each following \texttt{main\_dca} run is initialized with the result of the previous temperature.
While for the first temperature up to eight iterations are required until convergence is reached, three to six iterations are usually enough for the subsequent runs, given the temperature steps are sufficiently small.
Lowering the temperature is usually done with decreasing step size.

In the beginning only single-particle~(sp) functions, i.e. the cluster Green's function and self-energy, are computed.
When we reach the relevant temperature regime in which the system exhibits an instability, we additionally measure the required cluster two-particle Green's function in a separate two-particle~(tp) run, which only performs a single DCA or \dcaplus iteration.
Subsequent to the \texttt{main\_dca} runs, we separately analyze the output of all tp-runs with \texttt{main\_analysis}.
Listing~\ref{listing:cooldown} provides a schedule of jobs for a typical cooldown procedure.

\begin{lstlisting}[
basicstyle=\footnotesize\ttfamily,
caption={Workflow of a DCA or \dcaplus cooldown.},
captionpos=b,
label=listing:cooldown]
mpirun -np 8 ./main_dca ./T=1/input_sp.json     # Initial self-energy: "zero".
mpirun -np 8 ./main_dca ./T=0.5/input_sp.json   # Initial self-energy: T=1/dca_sp.hdf5.
mpirun -np 8 ./main_dca ./T=0.25/input_sp.json  # Initial self-energy: T=0.5/dca_sp.hdf5.

mpirun -np 8 ./main_dca ./T=0.1/input_sp.json   # Initial self-energy: T=0.25/dca_sp.hdf5.
mpirun -np 8 ./main_dca ./T=0.1/input_tp.json   # Initial self-energy: T=0.1/dca_sp.hdf5.
mpirun -np 8 ./main_dca ./T=0.09/input_sp.json  # Initial self-energy: T=0.1/dca_sp.hdf5.
mpirun -np 8 ./main_dca ./T=0.09/input_tp.json  # Initial self-energy: T=0.09/dca_sp.hdf5.
mpirun -np 8 ./main_dca ./T=0.08/input_sp.json  # Initial self-energy: T=0.09/dca_sp.hdf5.
mpirun -np 8 ./main_dca ./T=0.08/input_tp.json  # Initial self-energy: T=0.08/dca_sp.hdf5.

mpirun -np 8 ./main_analysis ./T=0.1/input_tp.json   # Reads T=0.1/dca_tp.hdf5.
mpirun -np 8 ./main_analysis ./T=0.09/input_tp.json  # Reads T=0.09/dca_tp.hdf5.
mpirun -np 8 ./main_analysis ./T=0.08/input_tp.json  # Reads T=0.08/dca_tp.hdf5.
\end{lstlisting}

We provide a python script that generates directories, input files and batch scripts for a cooldown.
By default, this script and the provided template input files are configured for DCA or \dcaplus calculation of the 2D single-band Hubbard model with on-site Coulomb interaction $U$ and fixed electron filling $d$, but it can easily be adjusted for other problems.
Usage of the script is explained in the \href{https://github.com/CompFUSE/DCA/wiki/Running}{Wiki}.

\section{Sustainable and scalable software development}
\label{sec:development}

The DCA++ project follows several well-proven methods in software engineering to develop and maintain a high-performance research code in a multidisciplinary team.
This section is devoted to sharing these methods and the tools that allow to implement them.
In addition to the dedicated article~\cite{Haehner:2018dca-development}, we want to call the interested reader's attention to the \emph{``How to'' documents} of the IDEAS Scientific Software Productivity Project~\cite{ideas}, which collect best practices in scientific software development, and which we partially follow here.

\subsection{Software testing and test-driven development}

Systematic testing is an essential element in software development to produce correct code.
In addition to detecting errors, software tests provide an important form of documentation by describing correct usage and defining requirements and limits with edge cases.

Software tests can be divided according to their level into \emph{unit}, \emph{integration}, and \emph{system-level tests}.
While unit tests operate at the smallest code scale testing single functions and classes, integration tests check the interaction between these units, and system-level tests validate and verify the behavior of the full code.
Orthogonal to this classification, we can categorize tests by their aspect into \emph{verification}, \emph{validation}, \emph{no-change}, and \emph{performance tests}~\cite{ideas}.
Verification tests determine whether the code satisfies specific requirements or conditions. They answer the question ``Are we building the product right?''~\cite{Boehm:1984is}.
Verification testing at the unit level is the tool kit of test-driven development (see below).
Validation tests, in contrast, are motivated by the question ``Are we building the right product?''~\cite{Boehm:1984is}.
These tests evaluate whether the software satisfies the intended purpose.
No-change tests define the current behavior of the software and are indispensable for refactoring legacy code.
Passing this type of software tests requires reproducing predefined results close to machine precision.
Performance tests are used to compare alternative algorithms, to monitor the code's performance, or to benchmark computing systems.

All of our software tests are automated and part of the regression test suite.
The regression test suite is divided with respect to required computational resources into \emph{fast tests} and \emph{extensive tests}, while performance tests are kept separated.
Fast tests include unit tests and lightweight integration tests.
Since they build and run fast, they can be executed frequently in the code development to detect problems early.
Extensive tests contain computationally more expensive integration tests and system-level tests, and are usually executed before a piece of code gets back into the main trunk (see workflow below).
They provide additional evidence that the code's behavior has not been unintentionally altered.

\emph{Test-driven development~(TDD)}~\cite{Beck:2002tdd} is a software development technique in which automated tests steer the design process.
Based on the two rules of only writing new code if there is a failing test and removing duplication, TDD consistently follows a three-step cycle:
\begin{enumerate}
    \item Create an automated test that fails.
    \item Make the test pass quickly by any means necessary.
    \item Refactor to eliminate duplication introduced by getting the test to pass.
\end{enumerate}
As any functional code is protected by at least one unit test, TDD prevents regression and delivers high confidence in the correctness of the software.
Writing tests first has the additional benefit that it encourages the programmer to think about the interface of the code first.
This way, TDD fosters designs that are high in cohesion and low in coupling.
The rhythm of TDD allows to work in small, safe steps and to focus on one task at a time.
The tests provide immediate feedback.

In the DCA++ project, we follow TDD whenever the existing code allows it and refactor code to make it more testable.
Where unit tests cannot be put in place, system-level no-change tests act as a global safety net.
Moreover, we require defects to be reported in the form of minimal, failing tests.
Our goal is a growing test suite and an increasing percentage of test driven code.

We implement tests using the Google C++ Testing Framework (Google Test)~\cite{googletest}.
Google Test builds on the xUnit architecture and provides benefits both at writing and executing tests.
It supports various fatal and non-fatal assertions, including death tests, and allows tests to share resources that are expensive to set up and not modified by the tests.
Moreover, tests can be value- or type-parameterized to cover a wide range of cases without duplicating test code.

\subsection{Software project management and collaborative development}

\paragraph*{Tools}
Successfully maintaining any software project begins with keeping the code base under version control.
By storing old versions of the software and monitoring changes to the code base, a version control system crucially contributes to reproducibility of simulation results and safe code development.
Version control systems that support branching allow the isolated implementation of features in dedicated branches, while the master branch, which contains the production code, is always operational.
We use the popular Git version control system~\cite{git}, in which each developer clones their own local repository.
Besides being distributed, Git offers fast and easy branching capabilities, which makes it well suited for collaborative software development.
Our Git repository is hosted on the GitHub platform, which, in addition to the repository, provides a set of tools for managing a software project enabling us to implement the workflow explained below.

While GitHub helps us to maintain the DCA++ source, we employ the EasyBuild framework to manage builds of DCA++ on production systems.
All dependencies of the DCA++ code, such as external libraries and compilers, but also the CMake build system, are specified in a build recipe.
Bundled in the so-called ``easyconfig'' file, the dependencies can be installed with a one-line command into a single module\footnote{More commonly, EasyBuild is used to not only build the dependencies but the full application or library.}.
This module can be loaded and used by other collaborators to easily build the DCA++ code.
We currently provide easyconfig files for CSCS' Piz Daint (hybrid Cray XC50 / XC40) and ORNL's Titan (Cray XK7) supercomputers.

To automate the quality control of the code, we run a Jenkins server~\cite{jenkins} hooked up to DCA++'s GitHub repository.
Triggered by changes to the repository, Jenkins builds all applications and tests, and runs the test suite.
A test report allows to immediately detect any degradation and to document the status of the code for later reference.

\begin{figure}[h]
    \centering
    \includegraphics[scale=0.3]{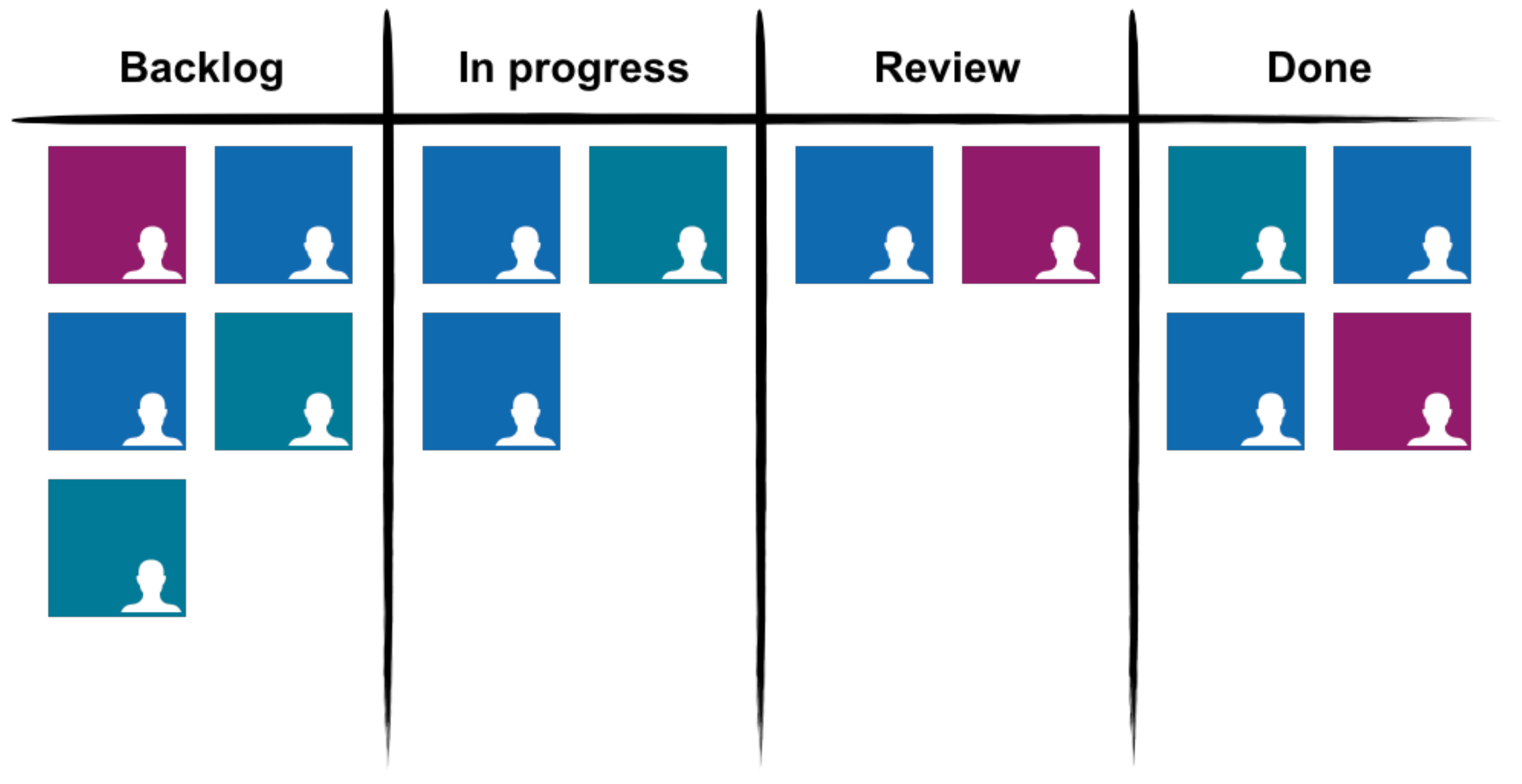}
    \caption{
        (Color online)
        Kanban board for software development. Each note corresponds to an issue.
        During its lifetime, an issue moves across the board from left to right.
    }
    \label{fig:kanban-board}
\end{figure}

\paragraph*{Workflow}
It is our goal to continuously and frequently deliver updates to the DCA++ code without compromising its quality.
To achieve this, we organize the code development by tracking each task, bug fix or other enhancement as an issue.
GitHub's issue tracker offers to categorize issues with labels and to assign them to a team member.
By associating issues with a milestone, the process to a specific goal can be monitored.
The implementation of a single issue should not exceed few weeks and more complex features, which we call \emph{epics}, should be broken down into smaller, independent tasks.
To visualize the status of all issues and to prioritize work, we use a Kanban board, a popular tool in agile development.
The board is organized in columns with each column representing a stage in the lifetime of an issue, i.e. \emph{Backlog}, \emph{In progress}, \emph{Review} and \emph{Done}.
A typical Kanban board for a software project is shown in Fig.~\ref{fig:kanban-board}.

\begin{figure}[h]
    \centering
    \includegraphics[scale=0.3]{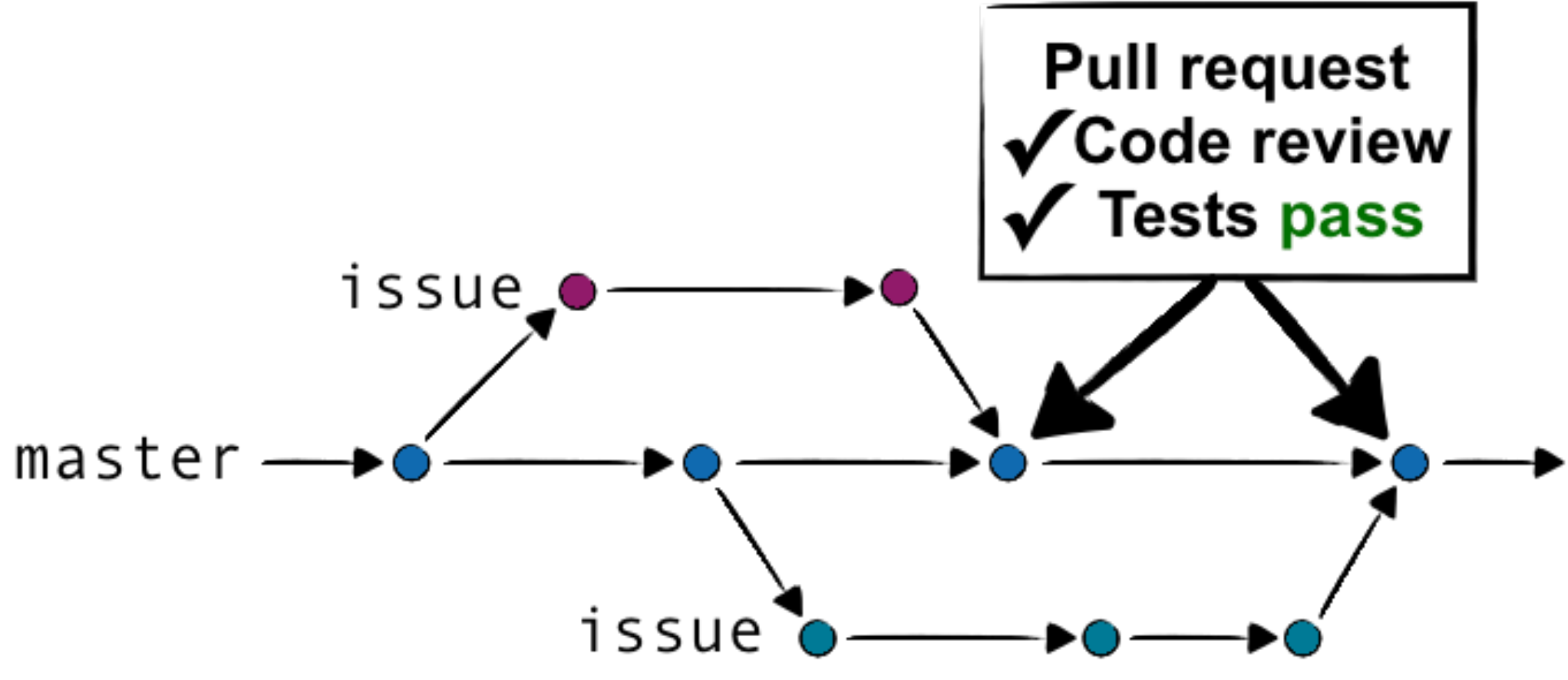}
    \caption{
        (Color online)
        Issue branch workflow.
        Each issue is implemented in a separate branch, while the master branch is constantly operational.
        Code reviews and automated testing at the end of an issue cycle ensure quality and correctness of the new code before it arrives in the master branch.
    }
    \label{fig:issue-branch-workflow}
\end{figure}

By developing each issue in a dedicated branch, we guarantee that the master branch is kept stable at all times.
As part of the quality control, the finished work has to be submitted in form of a pull request.
An experienced developer then reviews the code and requests changes if necessary.
If the reviewer approves the new code and Jenkins confirms that all tests pass, it will be merged into the master branch and become production code.
Fig.~\ref{fig:issue-branch-workflow} illustrates this \emph{issue branch workflow}.

For a successful realization of the workflow it is important to have early discussions about design decisions and implementation details.
This assures that the development of a new issue starts immediately in the right direction.
Collaborative development in large teams once more requires keeping individual issues smalls.
As part of continuous integration, we demand developers to rapidly check in their changes via pull requests to master.
This allows them to get feedback frequently and prevents a scenario, called \emph{merge hell}, in which it involves considerable effort to integrate a far diverged branch back into the main line.
Short issue cycles also mean that new features and bug fixes are quickly available to team members and users.
This can, in particular, avoid duplication of code work.

\section{Examples}
\label{sec:examples}

The following two standard use cases study superconductivity in the 2D Hubbard model.
The lattice model is described by the Hamiltonian
\begin{equation}
\label{eq:Hubbard}
    H = -t \sum_{\langle i, j \rangle, \sigma} c_{i \sigma}^\dagger c_{j \sigma} + U \sum_i n_{i \uparrow} n_{i \downarrow} \,,
\end{equation}
where $c_{i \sigma}^{\dagger}$ ($c_{i \sigma}$) creates (annihilates) an electron with spin $\sigma$ on lattice site $i$, $n_{i \sigma} = c_{i \sigma}^\dagger c_{i \sigma}$ is the corresponding number operator and $U$ denotes the on-site Coulomb repulsion.
In the form above, where $\langle i, j\rangle$ indicates that the first sum only runs over pairs of nearest neighbor sites, the model is restricted to nearest neighbor hopping with isotropic amplitude $t$.
The simulations are performed on a square lattice.

In the first example, we determine the superconducting transition temperature $T_c$ for a small $2\times2$ cluster employing the DCA method.
The second example provides a large-scale DCA$^+$ calculation in which we converge $T_c$ with respect to the cluster size $N_c$.
In both cases we use the CT-AUX QMC algorithm to solve the effective cluster problem.

\subsection{Computing the superconducting transition temperature $T_c$ with DCA}
\label{subsec:example_DCA}

\begin{figure}[h!]
    \centering
    \includegraphics[width=0.475\columnwidth]{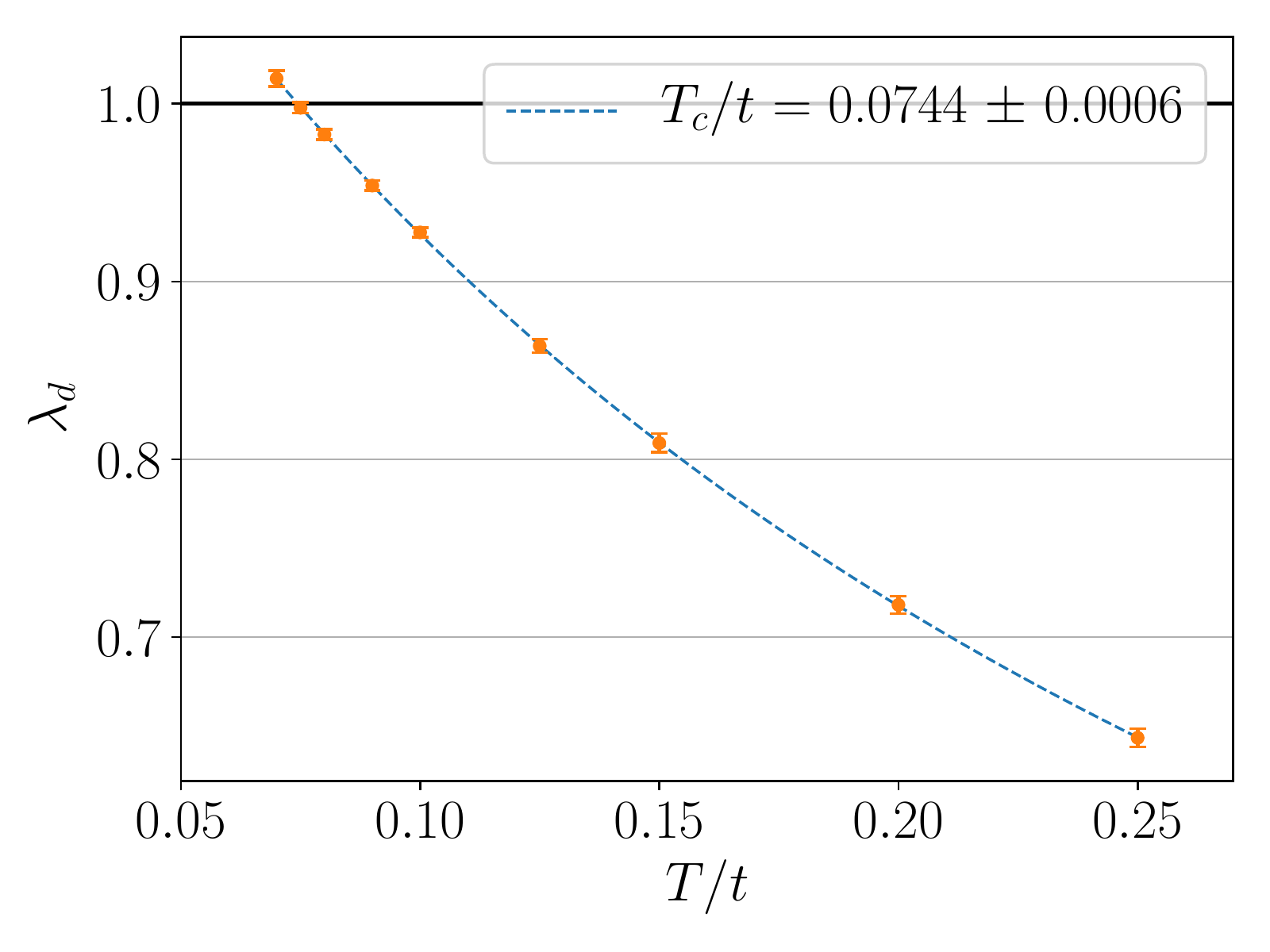}
    \caption{
        (Color online)
        Leading ($d$-wave) eigenvalue of the particle-particle Bethe-Salpeter kernel in the DCA calculated for $U = 8t$, $\langle n \rangle = 0.95$ and $N_c=4$.
        The intersection of $\lambda_d(T)$ with one yields the transition temperature $T_c$.
        Standard deviations illustrated as error bars were computed from 10 independent simulations that each performed 1~000~000 Monte Carlo measurements per iteration.
    }
    \label{fig:DCA_lambda_d-vs-T}
\end{figure}

As explained in section~\ref{subsec:susceptibilities}, we can analyze the transition to the superconducting phase by computing the leading eigenvalue of the Bethe-Salpeter equation in the particle-particle channel.
The superconducting state of the 2D Hubbard model is characterized by $d$-wave symmetry~\cite{Staar:2014gz, Maier:2005im}.
In Fig.~\ref{fig:DCA_lambda_d-vs-T} we plot the leading $d$-wave eigenvalue $\lambda_d$ as a function of temperature for strong coupling $U = 8t$, electron filling $\langle n \rangle = 0.95$ and $N_c = 4$.
As the temperature is lowered, the eigenvalue increases and the system approaches the phase instability.
After fitting the data points with a function of the form
\begin{equation}
    \lambda_d(T) = \frac{a}{\left( T-b \right)^c} \,,
\end{equation}
we can solve for the transition temperature $T_c$, defined as the temperature for which $\lambda_d(T)$ crosses one, i.e. $\lambda_d(T_c) = 1$.

\begin{figure}[h]
    \centering
    \begin{subfigure}{0.475\columnwidth}
        \centering
        \includegraphics[width=\columnwidth]{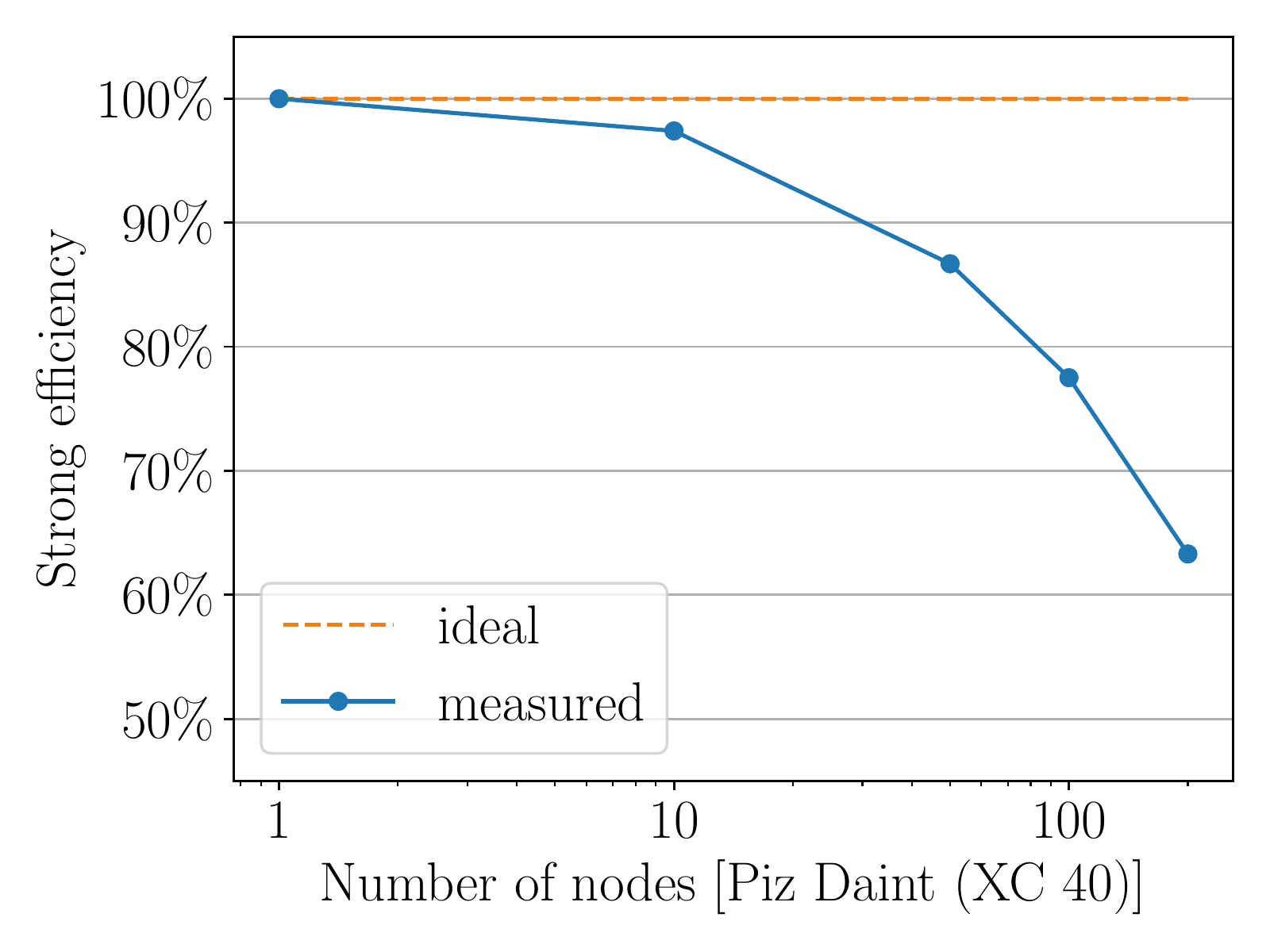}
    \end{subfigure}
    \hspace{0.03\columnwidth}
    \begin{subfigure}{0.475\columnwidth}
        \centering
        \includegraphics[width=\columnwidth]{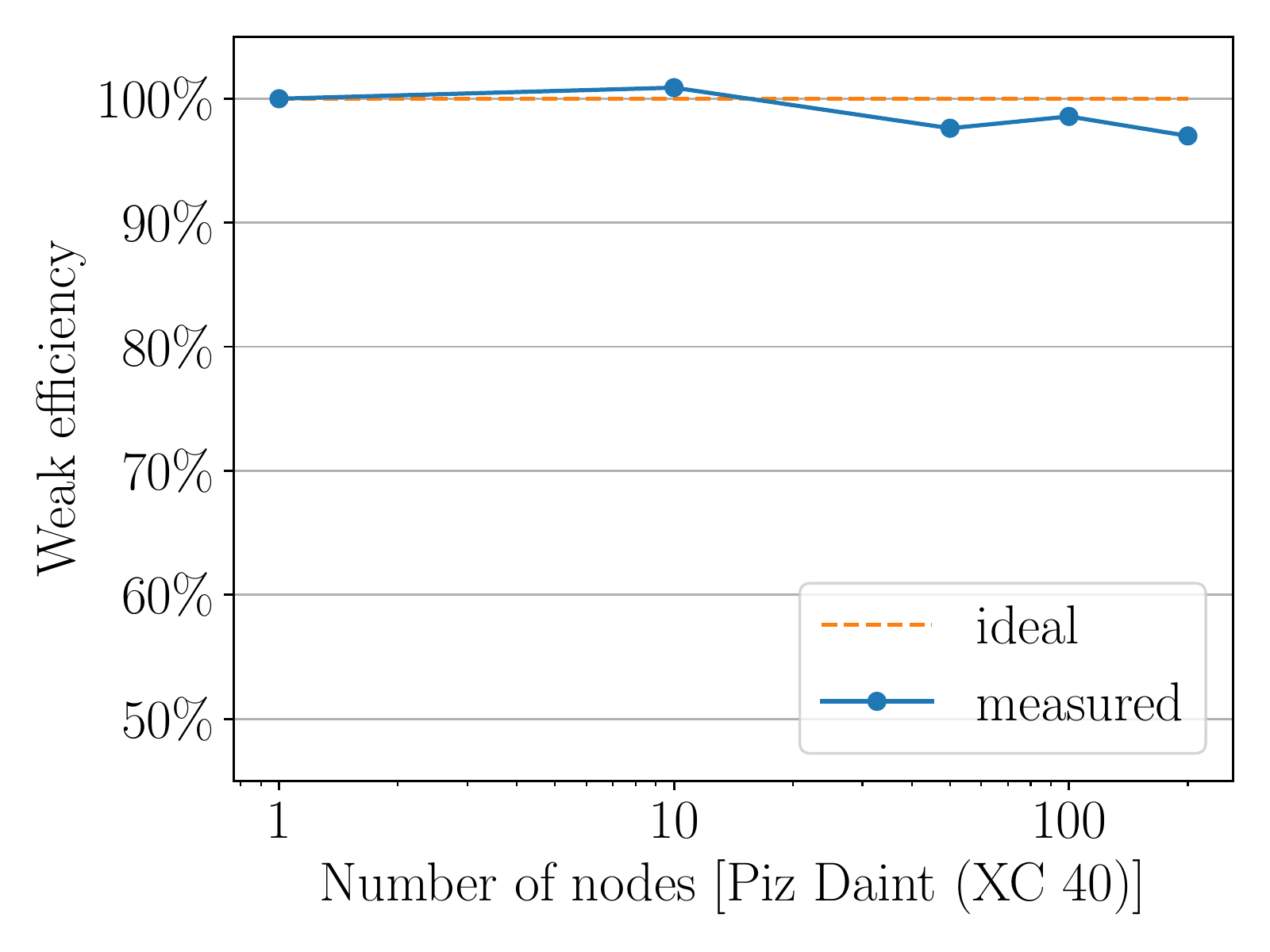}
    \end{subfigure}
    \caption{
        (Color online)
        Strong (left) and weak (right) scaling efficiency for 1 to 200 Cray XC40 multi-core nodes.
        Results represent time-to-solution of the full cooldown.
        We used 1~000~000 Monte Carlo measurements per iteration for the strong scaling analysis, while weak scaling is based on 1~000~000 Monte Carlo measurements per iteration and node.
    }
    \label{fig:mc-scaling}
\end{figure}

We performed the computation on Piz Daint's multi-core partition, a Cray XC40 system equipped with two 18-core Intel Xeon processors (2.1 GHz) per node.
Using a single compute node, results for 1~000~000 Monte Carlo measurements per iteration were obtained in 3.5 hours.
The time-to-solution~(TTS) can be greatly reduced by exploiting the code's parallel performance and increasing the number of nodes.
To quantify this, we measured the \emph{strong scaling} efficiency $E_s(n)$ as a function of the number of nodes $n$.
It is defined as
\begin{equation}
    E_s(n) = \frac{TTS(1)}{n \times TTS(n)} \times 100\% \,.
\end{equation}
Results are shown in the left plot of Fig.~\ref{fig:mc-scaling}.
Even though the computational demand of the problem is small, DCA++ allows to efficiently use up to 200 compute nodes, for which the TTS reduces to less than two minutes.

We also examined the \emph{weak scaling} behavior, where the problem size (number of Monte Carlo measurements) per node is kept constant.
Weak scaling efficiency $E_w(n)$ as a function of the number of nodes $n$ is computed as
\begin{equation}
    E_w(n) = \frac{TTS(1)}{TTS(n)} \times 100\% \,.
\end{equation}
The right plot of Fig.~\ref{fig:mc-scaling} presents the results for 1~000~000 Monte Carlo measurements per iteration and node.
While overall, we observe weak scaling efficiency close to the ideal case of 100\%, the 10 node result actually shows a deviation to ``better than ideal''.
This can be explained by the fact that we only kept the problem size with respect to the number of Monte Carlo measurements constant per node.
Other parts of the code, in particular the coarse-graining step, benefited from a speed-up when the number of nodes was increased.

\subsection{Cluster size scaling of $T_c$ with DCA$^+$}
\label{subsec:example_DCA+}

\begin{figure}[h]
    \centering
    \begin{minipage}[t]{0.475\columnwidth}
        \centering
        \includegraphics[width=\columnwidth]{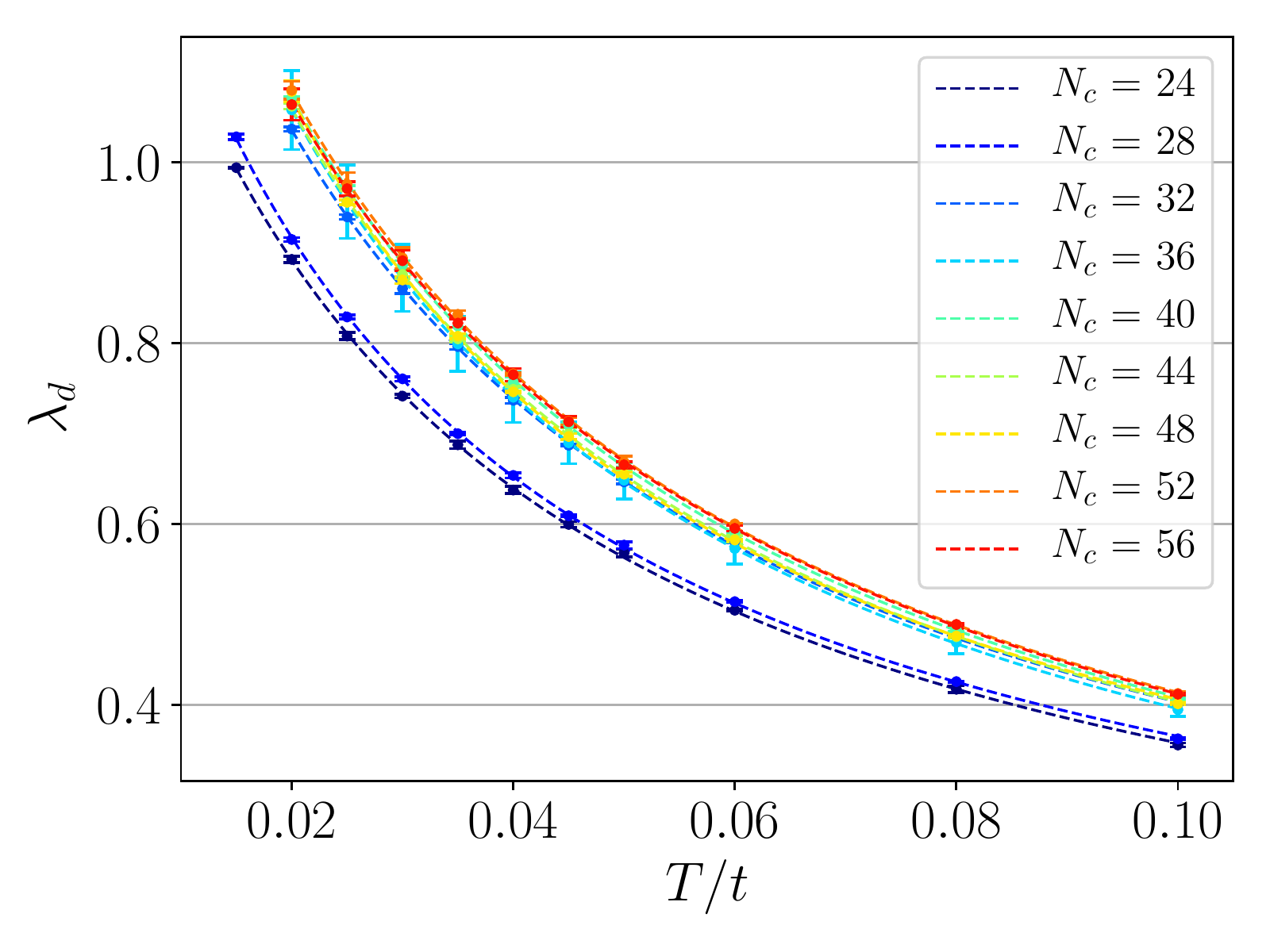}
        \caption{
            (Color online)
            Leading ($d$-wave) eigenvalue of the particle-particle Bethe-Salpeter kernel in the DCA$^+$ framework for $U = 4t$, $\langle n \rangle = 0.9$ and various cluster sizes.
            Error bars present, with the exception of $N_c = 36$, the standard deviation of four independent simulations with 1~000~000 Monte Carlo measurements each.
            For $N_c = 36$, the error bars depict the standard deviation of results for four different cluster shapes.
        }
        \label{fig:DCA+_lambda_d_vs_T}
    \end{minipage}
    \hspace{0.03\columnwidth}
    \begin{minipage}[t]{0.475\columnwidth}
        \centering
        \includegraphics[width=\columnwidth]{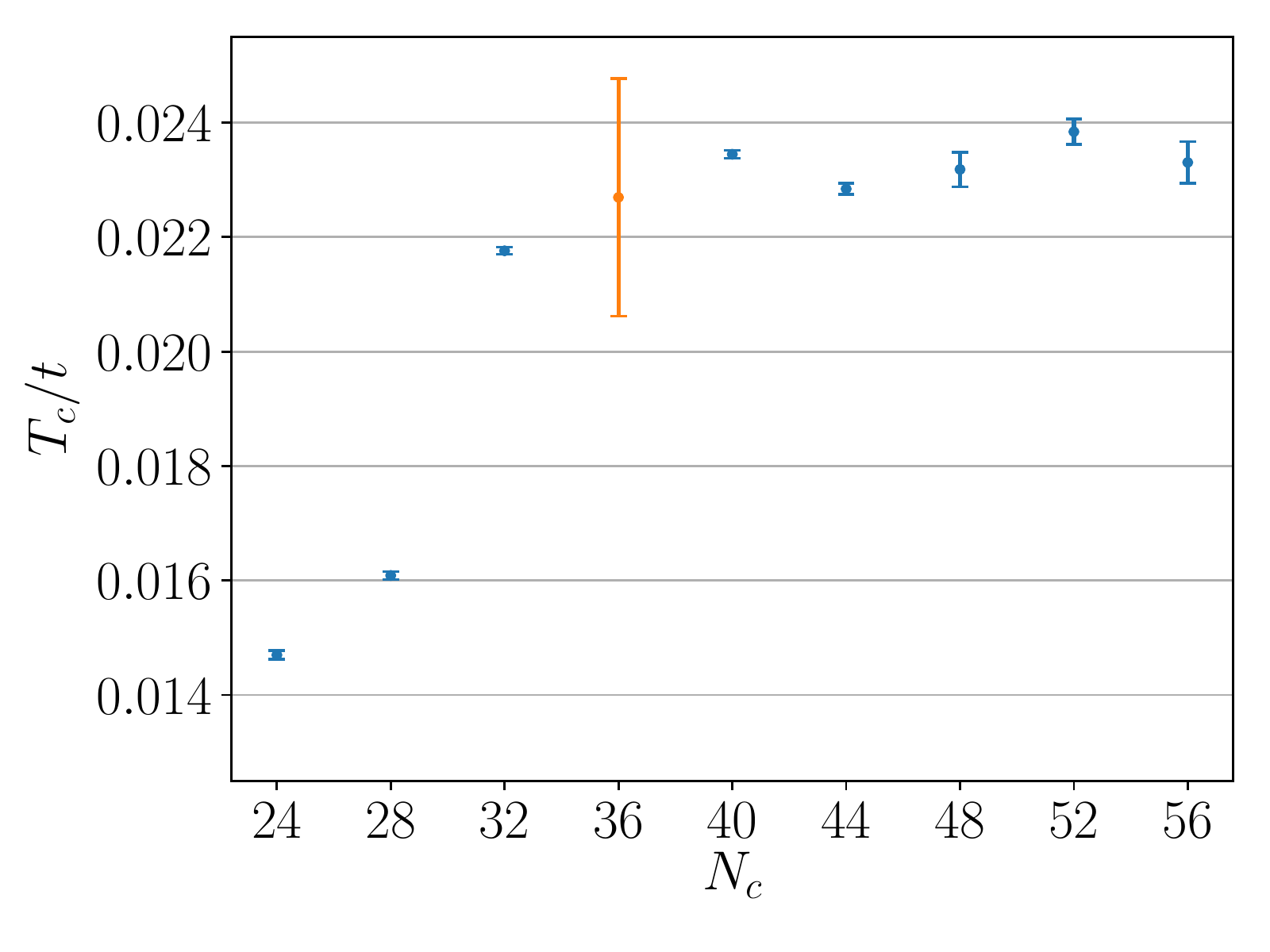}
        \caption{
            (Color online)
            DCA$^+$ results for the cluster size dependence of the transition temperature $T_c$ for $U = 4t$ and $\langle n \rangle = 0.9$.
            Error bars present, with the exception of $N_c = 36$, the standard deviation of four independent simulations with 1~000~000 Monte Carlo measurements each.
            For $N_c = 36$, the error bar depicts the standard deviation of results for four different cluster shapes.
        }
        \label{fig:DCA+_Tc_vs_Nc}
    \end{minipage}
\end{figure}

The reduced fermionic sign problem in the DCA$^+$ algorithm makes clusters sizes accessible that have been out of reach for standard DCA or finite-size QMC calculations.
This allows to enter a regime of asymptotic convergence, in which the transition temperature $T_c$ becomes independent of the cluster size.
Here, we consider the weak-coupling case of $U = 4t$ at 10\% doping.
Fig.~\ref{fig:DCA+_lambda_d_vs_T} shows the temperature dependence of the leading $d$-wave eigenvalue $\lambda_d(T)$ for cluster sizes between 24 and 56 and Fig.~\ref{fig:DCA+_Tc_vs_Nc} plots the resulting values for $T_c$ versus cluster size $N_c$.
We observe convergence of the transition temperature $T_c$ for clusters larger than 32 sites.

Error bars in Figs.~\ref{fig:DCA+_lambda_d_vs_T} and \ref{fig:DCA+_Tc_vs_Nc} represent, with the exception of $N_c = 36$, the statistical error of the QMC algorithm.
They tend to grow with the size of the cluster.
This originates from the behavior of the fermionic sign problem, which becomes more severe with increasing cluster size leading to a larger statistical error.

For the case of $N_c = 36$, we used four different cluster shapes to illustrate the error associated with the choice of cluster for given cluster size.
While the cluster shape dependence is substantially reduced by the introduction of \dcaplus~\cite{Staar:2013ec}, the corresponding spread in results still exceeds the statistical uncertainty of the Monte Carlo algorithm.
Ref.~\cite{Staar:2014gz} extends this analysis to more cluster sizes.

\begin{figure}[h!]
    \centering
    \begin{subfigure}[t]{0.475\columnwidth}
        \centering
        \includegraphics[width=\columnwidth]{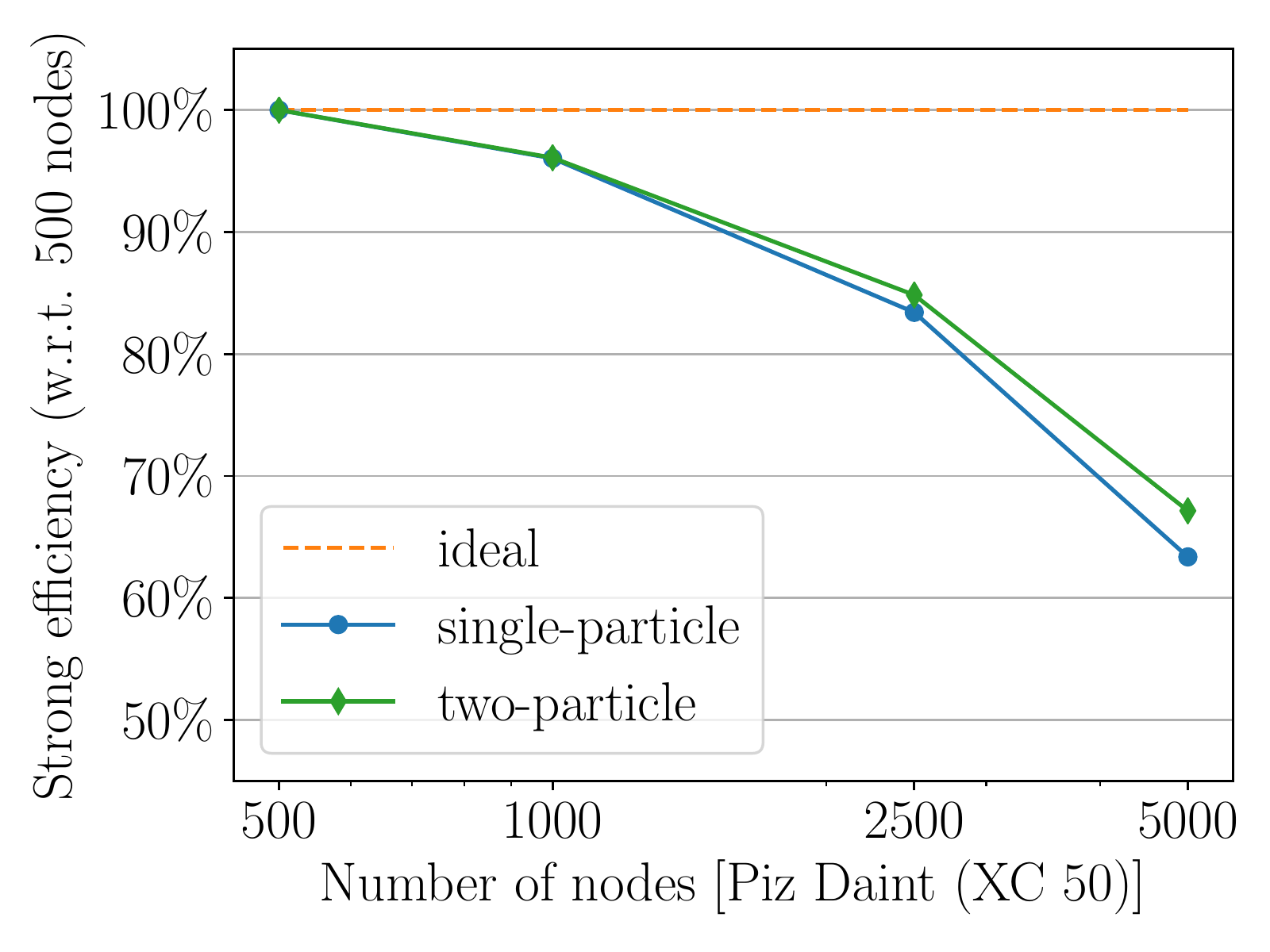}
    \end{subfigure}
    \hspace{0.03\columnwidth}
    \begin{subfigure}[t]{0.475\columnwidth}
        \centering
        \includegraphics[width=\columnwidth]{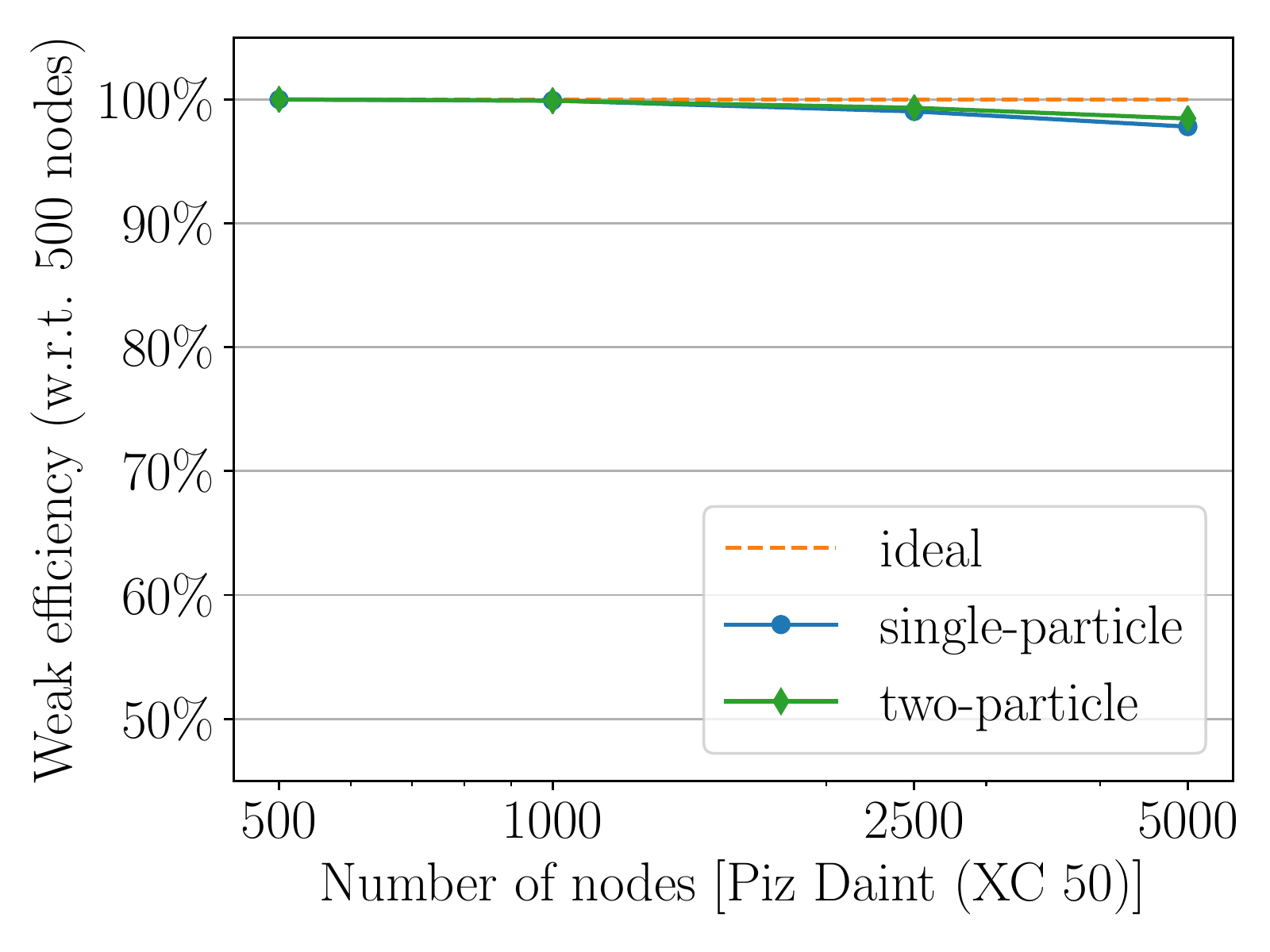}
    \end{subfigure}
    \caption{
        (Color online)
        Strong (left) and weak (right) scaling efficiency for 500 to 5000 Cray XC50 hybrid CPU-GPU nodes.
        Results represent time-to-solution of the QMC integration for $N_c = 56$ and $T = 0.02t$.
        Efficiencies are computed with respect to the 500 nodes result.
        We used 1~000~000 Monte Carlo measurements for the strong scaling analysis, while weak scaling is based on 2000 Monte Carlo measurements per node.
        Data distinguish between single-particle runs (blue dots), in which only single-particle quantities were accumulated, and two-particle runs (green diamonds), in which also the cluster two-particle Green's function was measured.
    }
    \label{fig:gpu-scaling}
\end{figure}

Large expansion orders resulting in large matrix operations make this problem computationally demanding, but also well suited for Piz Daint's Cray XC50 hybrid compute nodes.
They provide, in addition to a 12-core Intel Xeon processor (2.6 GHz), an NVIDIA Tesla P100 GPU accelerator.
This leadership computing system was used to examine the large-scale parallel performance of the DCA++ code.
We measured the TTS of the QMC integration for the computing intensive case of $N_c = 56$ and $T = 0.02t$.
Due to the problem size, we chose a minimum of 500 computes nodes to limit the run time.
For illustration, the full cooldown for $N_c = 56$ takes about 5.2 hours on 500 nodes.
We performed 1~000~000 Monte Carlo measurements for each node count to investigate the strong scaling behavior and used 2000 Monte Carlo measurements per node in the weak scaling analysis.
Fig.~\ref{fig:gpu-scaling} presents strong and weak scaling results for up to 5000 compute nodes.
The observed strong scaling efficiency implies that we achieve a considerable speed-up over 500 nodes when using nearly the entire machine.
As in the multi-core case, weak scaling efficiency excels reaching almost 100\%.

\begin{figure}[h!]
    \centering
    \includegraphics[width=0.475\columnwidth]{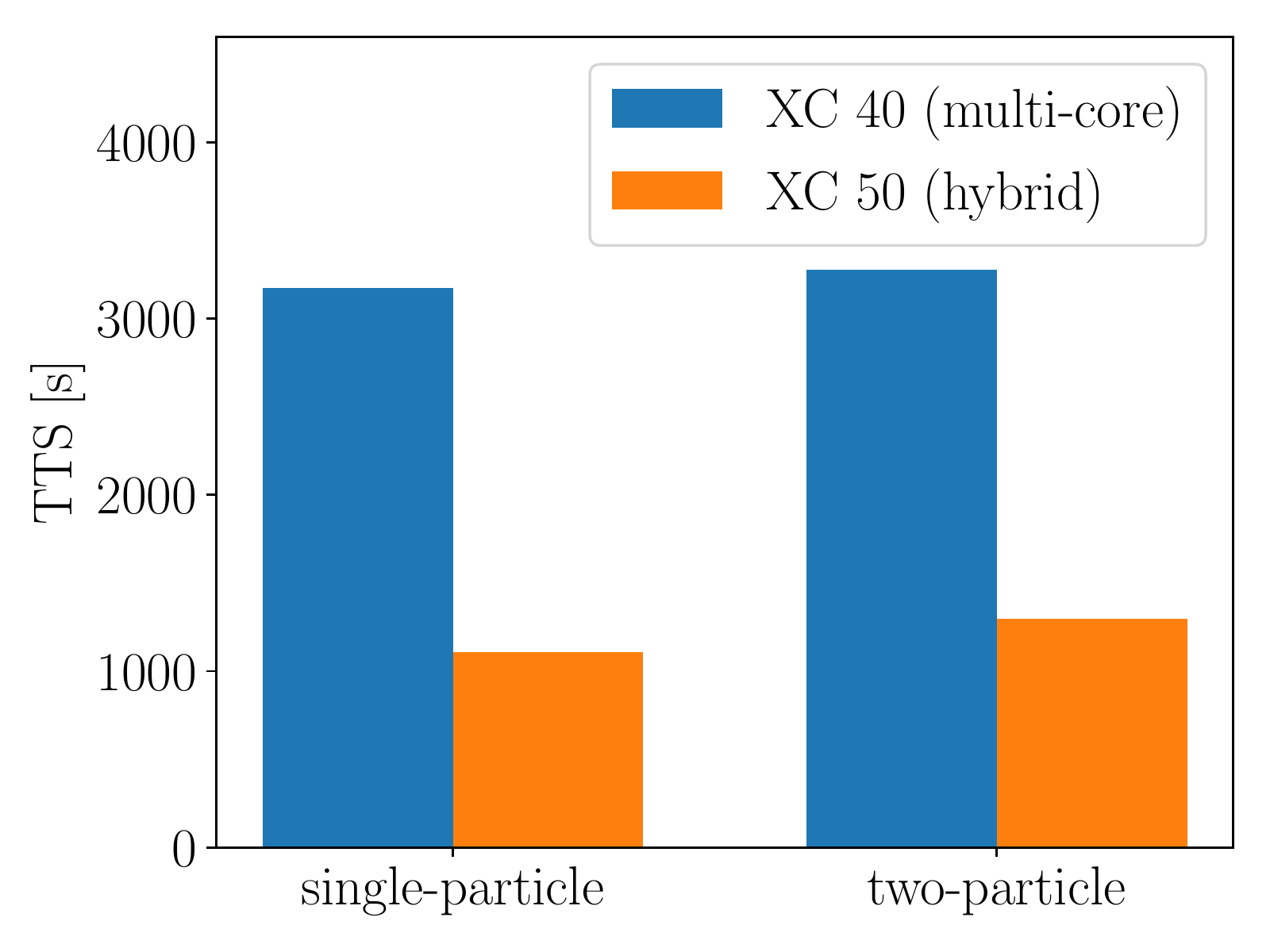}
    \caption{
        (Color online).
        Comparison of the time-to-solution of the QMC integration for $N_c = 56$, $T = 0.02t$ and 2000 Monte Carlo measurements between a Cray XC40 multi-core node and a Cray XC50 hybrid CPU-GPU node.
        Results are shown for a single-particle run (left), in which only single-particle quantities were accumulated, and a two-particle run (right), in which also the cluster two-particle Green's function was measured.
    }
    \label{fig:hybrid-vs-mc}
\end{figure}

We want to conclude this example by looking at the speed-up that a Cray XC50 hybrid node provides over a Cray XC40 multi-core node.
As in the scaling analysis, we compare the TTS of the QMC integration for the 56-site cluster at $T = 0.02t$.
The total number of Monte Carlo measurements per node was fixed to 2000.
We can see in Fig.~\ref{fig:hybrid-vs-mc} that the speed-up in a single-particle run is close to threefold, while the two-particle run is accelerated by a factor of 2.5.
The smaller speed-up when the computationally expensive cluster two-particle Green's function is accumulated accounts to the fact that the hybrid implementation of the CT-AUX solver performs the measurements on the host CPU.

\section{Final remarks}

\paragraph*{Future development}
DCA++ is an active software project with constantly extended functionality.
The next release will add full multi-orbital support and a state of the art implementation of the continuous-time interaction expansion QMC algorithm~(CT-INT)~\cite{Rubtsov:2004ib} to allow for simulations of multi-orbital models of systems such as the iron-based superconductors.

Since Monte Carlo algorithms are based on random sampling, they are non-deterministic by nature.
While implementations of these algorithms can still be tested at the unit level, it is often difficult to design conventional integration or system-level tests due to the random behavior.
For this reason, we have developed a framework based on hypothesis testing, which is able to validate stochastic programs.
This framework is currently being integrated into DCA++.

The workload balance in the CT-AUX solver on hybrid CPU-GPU systems, i.e. running the Monte Carlo walker on the GPU and accumulating measurements on the CPU, has proven to be efficient on current systems.
However, to respond to the emerging GPU-dominated node configurations, the next release of DCA++ will provide the capability to move the CT-AUX accumulator to the GPU, too.
In the long term, the growing diversity of HPC architectures will make it necessary to revisit the on-node parallelization strategy altogether.
We plan to follow a task based parallelization model based on the HPX-3 system~\cite{Kaiser:2014fy} to ensure portability and efficiency for the future.

\paragraph*{Getting started}
The DCA++ code is licensed under the BSD 3-Clause License.
Hence, it is open source and a version of the code can easily be obtained from the public GitHub repository at \url{https://github.com/CompFUSE/DCA}.
We advise prospective users to visit the \href{https://github.com/CompFUSE/DCA/wiki}{Wiki} section of the repository, which, in particular, provides a small \href{https://github.com/CompFUSE/DCA/wiki/Tutorials}{tutorial}.
Questions about building or running DCA++ can be posted to the repository's built-in \href{https://github.com/CompFUSE/DCA/issues}{issue tracking system}.

\paragraph*{Contributing back}
There are various ways to contribute to the DCA++ project.
Reports about bugs and other problems can be submitted via the \href{https://github.com/CompFUSE/DCA/issues}{issue tracking system}.
The issue tracker is also the preferred communication channel for any other request concerning the code, including the suggestion of new features.
We welcome code contributions, too.
They are best submitted by creating a \href{https://github.com/CompFUSE/DCA/pulls}{pull request} in the GitHub repository.

\paragraph*{Citation guideline}
If the DCA++ code contributes to your research, we kindly ask to acknowledge this in any resulting scientific publication by citing this paper.

\section*{Acknowledgments}

The authors would like to acknowledge the contributions of Giovanni Balduzzi, Peter W. Doak, Mi Jiang, Andrei Plamada and Bart Ydens to the DCA++ project.
We would like to thank John Biddiscombe for his contributions to refactoring the code base, Guilherme Peretti-Pezzi for his assistance in setting up continuous integration and his support with EasyBuild, and Philipp Werner for providing the CT-HYB QMC solver.
U.R.H. acknowledges the support by the National Center of Competence in Research (NCCR) MARVEL, funded by the Swiss National Science Foundation.
The work of T.A.M. was supported by the Scientific Discovery through Advanced Computing~(SciDAC) program funded by U.S. Department of Energy, Office of Science, Advanced Scientific Computing Research and Basic Energy Sciences, Division of Materials Sciences and Engineering.
This research used resources of the Oak Ridge Leadership Computing Facility~(OLCF) awarded by the INCITE program, and of the Swiss National Supercomputing Center~(CSCS).
OLCF is a DOE Office of Science User Facility supported under Contract DE-AC05-00OR22725.

\appendix

\section{CMake options}
\label{app:cmake_options}

The DCA++ build can be configured with several CMake options.
These options can either be appended to the \texttt{cmake} command with the \texttt{-D} flag:

\begin{lstlisting}[frame=]
$ cmake ../dca_source -D<variable1>=<value1> -D<variable2>=<value2> ...
\end{lstlisting}

\noindent or specified using the \texttt{ccmake} interface, which is launched in the existing build directory (after \texttt{cmake} was called) with:

\begin{lstlisting}[frame=]
$ ccmake .
\end{lstlisting}

\noindent Table~\ref{tab:cmake_options} lists some general and all DCA++ specific CMake options.

\newpage
\begin{table}[h!]
    \centering
    \caption{Relevant CMake options for building DCA++.}
    \label{tab:cmake_options}
    \footnotesize
    \begin{tabularx}{\textwidth}{l l X}
        \hline \hline
        Option & Default value & Description \\ \hline
        \texttt{CMAKE\_BUILD\_TYPE} & & The build type. Options are: \newline [empty], \texttt{Debug}, \texttt{Release}, \texttt{RelWithDebInfo}, \texttt{MinSizeRel}. According to the build type CMake sets different optimization and debug compiler flags. \texttt{Release} is recommended for production runs.\\
        \hline
        \texttt{FFTW\_INCLUDE\_DIR} & & Path to \texttt{fftw3.h}. \\
        \texttt{FFTW\_LIBRARY} & & The FFTW3(-compatible) library. If the \texttt{-D} flag is used, multiple libraries have to be separated by semicolons and enclosed by double quotes. \\
        \hline
        \texttt{DCA\_HAVE\_LAPACK} & & Set to \texttt{TRUE} if you use a compiler wrapper that automatically links against BLAS and LAPACK libraries, to prevent CMake from searching for these libraries. \\
        \hline
        \texttt{DCA\_WITH\_MPI} & \texttt{ON} & Enable MPI. If MPI is enabled, set the environment variable \texttt{CXX} to the MPI compiler wrapper before running CMake. \\
        \hline
        \texttt{DCA\_WITH\_CUDA} & \texttt{OFF} & Enable GPU support. \\
        \texttt{DCA\_WITH\_PINNED\_HOST\_MEMORY} & \texttt{OFF} & Enable pinned host memory. \emph{(advanced)} \\
        \texttt{CUDA\_GPU\_ARCH} & \texttt{sm\_60} & Name of the \emph{real} architecture to build for. \\
        \texttt{CUDA\_TOOLKIT\_ROOT\_DIR} & & Path to the CUDA Toolkit. Determined by CMake. Set it manually, if CMake cannot find it. \\
        \texttt{MAGMA\_DIR} & & Path to the MAGMA installation directory. Hint for CMake to find MAGMA. \\
        \hline
        \texttt{DCA\_BUILD\_DCA} & \texttt{ON} & Build \texttt{main\_dca}. \\
        \texttt{DCA\_BUILD\_ANALYSIS} & \texttt{ON} & Build \texttt{main\_analysis}. \\
        \texttt{DCA\_BUILD\_CLUSTER\_SOLVER\_CHECK} & \texttt{OFF} & Build \texttt{cluster\_solver\_check}, an application that tests a QMC solver against exact diagonalization. \\
        \hline
        \texttt{DCA\_WITH\_TESTS\_FAST} & \texttt{OFF} & Build DCA++'s fast tests. \\
        \texttt{DCA\_WITH\_TESTS\_EXTENSIVE} & \texttt{OFF} & Build DCA++'s extensive tests. \\
        \texttt{DCA\_WITH\_TESTS\_PERFORMANCE} & \texttt{OFF} & Build DCA++'s performance tests. (Only in \texttt{Release} mode.)\\
        \texttt{TEST\_RUNNER} & & Command for executing (MPI) programs. Required if MPI is enabled or on systems that use a command for launching executables (e.g. \texttt{aprun} or \texttt{srun}). \\
        \texttt{MPIEXEC\_NUMPROC\_FLAG} & \texttt{-n} & Flag used by \texttt{TEST\_RUNNER} to specify the number of processes. \\
        \texttt{MPIEXEC\_PREFLAGS} & & Flags to pass to \texttt{TEST\_RUNNER} directly before the executable to run. \\
        \hline
        \texttt{DCA\_CLUSTER\_SOLVER} & \texttt{CT-AUX} & The cluster solver for the DCA/DCA$^+$ calculation. Options are: \texttt{CT-AUX}, \texttt{SS-CT-HYB}. \\
        \texttt{DCA\_WITH\_THREADED\_SOLVER} & \texttt{ON} & Use multiple walker and accumulator threads in the cluster solver. \\
        \texttt{DCA\_LATTICE} & \texttt{square} & Lattice type, options are: \texttt{bilayer}, \texttt{square}, \texttt{triangular}. \\
        \texttt{DCA\_POINT\_GROUP} & \texttt{D4} & Point group symmetry, options are: \texttt{C6}, \texttt{D4}. \\
        \texttt{DCA\_MODEL} & \texttt{tight-binding} & Model type, options are: \texttt{tight-binding}. \\
        \texttt{DCA\_RNG} & \texttt{std::mt19937\_64} & Random number generator, options are: \texttt{std::mt19937\_64}, \texttt{std::ranlux48}, \texttt{custom}. See Random number generator section in the Wiki for details. \\
        \texttt{DCA\_RNG\_CLASS} &  & Class name including namespaces of the \emph{custom} random number generator. \\
        \texttt{DCA\_RNG\_HEADER} & & Header file of the \emph{custom} random number generator. \\
        \texttt{DCA\_RNG\_LIBRARY} & & \emph{Custom} random number generator library. \\
        \texttt{DCA\_PROFILER} & \texttt{None} & Profiler type, options are: \texttt{None}, \texttt{Counting}, \texttt{PAPI}. \\
        \texttt{DCA\_WITH\_AUTOTUNING} & \texttt{OFF} & Enable auto-tuning. Requires a profiler type other than \texttt{None}. \emph{(advanced)} \\
        \texttt{DCA\_WITH\_GNUPLOT} & \texttt{OFF} & Enable Gnuplot. \\
        \texttt{DCA\_WITH\_SINGLE\_PRECISION\_MEASUREMENTS} & \texttt{OFF} & Measure in single precision. \\
        \texttt{DCA\_WITH\_SINGLE\_PRECISION\_COARSEGRAINING} & \texttt{OFF} & Coarsegrain in single precision. \emph{(advanced)} \\
        \texttt{DCA\_WITH\_QMC\_BIT} & \texttt{OFF} & Enable QMC solver built-in tests. \emph{(advanced)} \\
        \hline \hline
    \end{tabularx}
\end{table}
\newpage

\section{Input parameters}
\label{app:parameters}

This appendix section provides short descriptions of each input parameter using the format:

\begin{itemize}
    \item \texttt{"parameter-name"}: type (\texttt{default value}) \\
    Description.
\end{itemize}

In addition, Listing~\ref{listing:input_file} shows a sample input file as used for the DCA$^+$ calculation of Section~\ref{subsec:example_DCA+}.

\subsection*{Output parameters}
{\footnotesize
Group \texttt{"output"}:

\begin{itemize}
    \item \texttt{"directory"}: string (\texttt{"./"}) \\
    Directory to write the output to.

    \item \texttt{"output-format"}: string (\texttt{"HDF5"}) \\
    File format of the output files.
    Options are: \texttt{"HDF5"}, \texttt{"JSON"}.

    \item \texttt{"filename-dca"}: string (\texttt{"dca.hdf5"}) \\
    Filename for the output of the application \texttt{main\_dca}.

    \item \texttt{"filename-analysis"}: string (\texttt{"analysis.hdf5"}) \\
    Filename for the output of the application \texttt{main\_analysis}.

    \item \texttt{"filename-ed"}: string (\texttt{"ed.hdf5"})  \\
    Filename for the ED output in the application \texttt{cluster\_solver\_check}.

    \item \texttt{"filename-qmc"}: string (\texttt{"qmc.hdf5"}) \\
    Filename for the QMC output in the application \texttt{cluster\_solver\_check}.

    \item \texttt{"filename-profiling"}: string (\texttt{"profiling.json"}) \\
    Filename for the profiling output.
    The file format is always JSON.

    \item \texttt{"dump-lattice-self-energy"}: boolean (\texttt{false}) \\
    Write out the lattice self-energy $\Sigma(\bk, \wn)$ in the application \texttt{main\_dca}.

    \item \texttt{"dump-cluster-Greens-functions"}: boolean (\texttt{false}) \\
    Write out various cluster Green's functions in the application \texttt{main\_dca}.

    \item \texttt{"dump-Gamma-lattice"}: boolean (\texttt{false}) \\
    Write out the lattice irreducible vertex function $\Gamma(k, k', q)$ in the application \texttt{main\_analysis}.

    \item \texttt{"dump-chi-0-lattice"}: boolean (\texttt{false}) \\
    Write out $\chi_0(k)$ in the application \texttt{main\_analysis}.
\end{itemize}}

\subsection*{Physics parameters}
{\footnotesize
Group \texttt{"physics"}:

\begin{itemize}
    \item \texttt{"beta"}: double (\texttt{1.}) \\
    Inverse temperature.

    \item \texttt{"density"}: double (\texttt{1.}) \\
    Target electron density.

    \item \texttt{"chemical-potential"}: double (\texttt{0.}) \\
    Initial value of the chemical potential.
    This value is overwritten, if an output file with an initial self-energy is read (see section DCA parameters).

    \item \texttt{"adjust-chemical-potential"}: boolean (\texttt{true}) \\
    Adjust the chemical potential to obtain the specified density.
\end{itemize}}

\subsection*{Model parameters}
\begin{itemize}[label=$\diamond$]

\item \emph{Single-band Hubbard model} \\
{\footnotesize
Used if \texttt{square} lattice or \texttt{triangular} lattice are selected. \\[5pt]
Group \texttt{"single-band-Hubbard-model"}:

\begin{itemize}[label=$\bullet$, leftmargin=*]
    \item \texttt{"t"}: double (\texttt{0.}) \\
    Nearest neighbor hopping parameter.

    \item \texttt{"t-prime"}: double (\texttt{0.}) \\
    Next nearest neighbor hopping parameter.

    \item \texttt{"U"}: double (\texttt{0.}) \\
    On-site Coulomb repulsion.

    \item \texttt{"V"}: double (\texttt{0.}) \\
    Nearest neighbor coulomb repulsion for opposite spins.

    \item \texttt{"V-prime"}: double (\texttt{0.}) \\
    Nearest neighbor coulomb repulsion for same spins.
\end{itemize}}

\item \emph{Bilayer Hubbard model} \\
{\footnotesize
Used if \texttt{bilayer} lattice is selected. \\[5pt]
Group \texttt{"bilayer-Hubbard-model"}:

\begin{itemize}[label=$\bullet$, leftmargin=*]
    \item \texttt{"t"}: double (\texttt{0.}) \\
    Nearest neighbor hopping parameter.

    \item \texttt{"t-prime"}: double (\texttt{0.}) \\
    Next nearest neighbor hopping parameter.

    \item \texttt{"t-perp"}: double (\texttt{0.}) \\
    Hopping parameter between layers.

    \item \texttt{"U"}: double (\texttt{0.}) \\
    On-site Coulomb repulsion.

    \item \texttt{"V"}: double (\texttt{0.}) \\
    Coulomb repulsion between layers for opposite spins.

    \item \texttt{"V-prime"}: double (\texttt{0.}) \\
    Coulomb repulsion between layers for same spins.
\end{itemize}}

\item \emph{Material model} \\
{\footnotesize
Group \texttt{"material-model"}:

\begin{itemize}[label=$\bullet$, leftmargin=*]
    \item \texttt{"t\_ij-filename"}: string (\texttt{"t\_ij.txt"}) \\
    Name of the CSV file containing the hopping matrix $t_{ij}$.

    \item \texttt{"U\_ij-filename"}: string (\texttt{"U\_ij.txt"}) \\
    Name of the CSV file containing the Coulomb repulsion matrix $U_{ij}$.
\end{itemize}}

\end{itemize}

\subsection*{DCA parameters}
{\footnotesize
Group \texttt{"DCA"}:

\begin{itemize}
    \item \texttt{"initial-self-energy"}: string (\texttt{"zero"}) \\
    Either the name of the file (including path) with the initial self-energy (usually the \texttt{main\_dca} output file of the previous temperature) or \texttt{"zero"} indicating that the initial self-energy should be zero.

    \item \texttt{"iterations"}: integer (\texttt{1}) \\
    Number of DCA/DCA$^+$ iterations.

    \item \texttt{"accuracy"}: double (\texttt{0.}) \\
    Stop the DCA/DCA$^+$ loop if this accuracy has been reached.

    \item \texttt{"self-energy-mixing-factor"}: double (\texttt{1.}) \\
    Parameter $\alpha$ of Eq.~(\ref{eq:self-energy-mixing}).

    \item \texttt{"interacting-orbitals"}: array of integers (\texttt{[0]}) \\
    Indices of orbitals that are treated interacting orbitals.
    This parameter must be consistent with the model that is used.

    \item \texttt{"do-finite-size-QMC"}: boolean (\texttt{false}) \\
    Do a finite-size QMC calculation (no mean-field).

\item Subgroup \texttt{"coarse-graining"}:

\begin{itemize}[label=$\bullet$]
    \item \texttt{"k-mesh-recursion"}: integer (\texttt{0}) \\
    Number of recursion steps in the creation of the momentum space mesh.
    See Ref.~\cite{Staar:2016ia} for details.

    \item \texttt{"periods"}: integer (\texttt{0}) \\
    Number of periods of the interlaced coarse-graining patches, i.e. degree of their interleaving.
    Restricted by \texttt{k-mesh-recursion}.
    See Ref.~\cite{Staar:2016ia} for details.

    \item \texttt{"quadrature-rule"}: integer (\texttt{1}) \\
    Determines the quadrature rule used in the coarse-graining. \\
    \texttt{quadrature-rule} $<$ 0: use a flat mesh. \\
    \texttt{quadrature-rule} $>=$ 0: use the Grundmann-Moeller rule~\cite{simplex_gm_rule} of index = \texttt{quadrature-rule}. \\
    A rule of index $s$ is exact for polynomials up to order $2s+1$.

    \item \texttt{"threads"}: integer (\texttt{1}) \\
    Number of threads used in the coarse-graining.

    \item \texttt{"tail-frequencies"}: integer (\texttt{0}) \\
    Number of tail frequencies used for updating the chemical potential.
\end{itemize}

\item Subgroup \texttt{"DCA+"}:

\begin{itemize}[label=$\bullet$]
    \item \texttt{"do-DCA+"}: boolean (\texttt{false}) \\
    Use the DCA$^+$ algorithm.

    \item \texttt{"deconvolution-iterations"}: integer (\texttt{16}) \\
    Maximum number of iterations in the deconvolution step.

    \item \texttt{"deconvolution-tolerance"}: double (\texttt{1.e-3}) \\
    Termination criteria for the deconvolution step.

    \item \texttt{"HTS-approximation"}: boolean (\texttt{false}) \\
    Use a high-temperature series approximation in the lattice mapping.

    \item \texttt{"HTS-threads"}: integer (\texttt{1}) \\
    Number of threads used in the HTS-solver.
\end{itemize}

\end{itemize}}

\subsection*{Domains parameters}
{\footnotesize
Group \texttt{"domains"}:

\begin{itemize}
\item Subgroup \texttt{"real-space-grids}:

\begin{itemize}[label=$\bullet$]
    \item \texttt{"cluster"}: array of arrays of integers (lattice basis, e.g. in 2D: \texttt{[[1, 0], [0, 1]]}) \\
    Real space DCA cluster.
    Given as coordinates with respect to the lattice basis. \\
    To choose a cluster of a given size, consult the exhaustive list of 2D and 3D clusters in \\
    \texttt{tools/cluster\_definitions.txt}.

    \item \texttt{"sp-host"}: array of arrays of integers (lattice basis, e.g. in 2D: \texttt{[[1, 0], [0, 1]]}) \\
    Real space host grid for single-particle functions, also called \emph{(sp-)lattice}.
    Given as coordinates with respect to the lattice basis.

    \item \texttt{"tp-host"}: array of arrays of integers (lattice basis, e.g. in 2D: \texttt{[[1, 0], [0, 1]]}) \\
    Real space host grid for two-particle functions, also called \emph{tp-lattice}.
    Only used in DCA$^+$. \\
    Given as coordinates with respect to the lattice basis.
\end{itemize}

\item Subgroup \texttt{"imaginary-time"}:

\begin{itemize}[label=$\bullet$]
    \item \texttt{"sp-time-intervals"}: integer (\texttt{128}) \\
    Discretization of the imaginary time axis.

    \item \texttt{"time-intervals-for-time-measurements"}: integer (\texttt{1}) \\
    Discretization of the imaginary time axis for additional time measurements.
\end{itemize}

\item Subgroup \texttt{"imaginary-frequency"}:

\begin{itemize}[label=$\bullet$]
    \item \texttt{"sp-fermionic-frequencies"}: integer (\texttt{256}) \\
    Number of fermionic Matsubara frequencies for single-particle functions.

    \item \texttt{"HTS-bosonic-frequencies"}: integer (\texttt{0}) \\
    Number of bosonic Matsubara frequencies in the HTS-solver.

    \item \texttt{"four-point-fermionic-frequencies"}: integer (\texttt{1}) \\
    Number of fermionic Matsubara frequencies for four-point functions.
\end{itemize}

\item Subgroup \texttt{"real-frequency"}:

\begin{itemize}[label=$\bullet$]
    \item \texttt{"min"}: double (\texttt{-10.}) \\
    Minimum real frequency.

    \item \texttt{"max"}: double (\texttt{10.}) \\
    Maximum real frequency.

    \item \texttt{"frequencies"}: integer (\texttt{3}) \\
    Number of real frequencies.

    \item \texttt{"imaginary-damping"}: double (\texttt{0.01}) \\
    Small positive shift $\eta$, used to shift the real frequencies $\omega$ away from the real axis, $\omega \rightarrow \omega + i \eta$.
\end{itemize}

\end{itemize}}

\subsection*{Monte Carlo integration parameters}
{\footnotesize
Group \texttt{"Monte-Carlo-integration"}:

\begin{itemize}
    \item \texttt{"seed"}: integer or string (\texttt{985456376}) \\
    Seed for the random number generator(s) used in the Monte Carlo integration.
    Instead of an integer, the string \texttt{"random"} can be passed to generate a \emph{random} seed.

    \item \texttt{"warm-up-sweeps"}: integer (\texttt{20}) \\
    Number of warm-up sweeps.

    \item \texttt{"sweeps-per-measurement"}: double (\texttt{1.}) \\
    Number of sweeps per measurement.

    \item \texttt{"measurements-per-process-and-accumulator"}: integer (\texttt{100}) \\
    Number of independent measurements each accumulator of each process performs.

    \item Subgroup \texttt{"threaded-solver"}:

    \begin{itemize}[label=$\bullet$]
        \item \texttt{"walkers"}: integer (\texttt{1}) \\
        Number of Monte Carlo walkers.

        \item \texttt{"accumulators"}: integer (\texttt{1}) \\
        Number of Monte Carlo accumulators.
    \end{itemize}

\end{itemize}}

\subsection*{Monte Carlo solver parameters}
\begin{itemize}[label=$\diamond$]

\item \emph{CT-AUX} \\
{\footnotesize
Used if \texttt{CT-AUX} is selected as the cluster solver.  \\[5pt]
Group \texttt{"CT-AUX"}:

\begin{itemize}[label=$\bullet$, leftmargin=*]
    \item \texttt{"expansion-parameter-K"}: double (\texttt{1.}) \\
    The perturbation order in the CT-AUX algorithm increases linearly with the expansion parameter $K$.
    While $K$ is only subject to the restriction of being positive, values of the order of one have proven to be a good choice~\cite{Gull:2008cm}.

    \item \texttt{"initial-configuration-size"}: integer (\texttt{10}) \\
    The CT-AUX solver is initialized with \texttt{initial-configuration-size} random \emph{interacting} vertices.

    \item \texttt{"initial-matrix-size"}: integer (\texttt{128}) \\
    Initial size of the CT-AUX matrices.

    \item \texttt{"max-submatrix-size"}: integer (\texttt{128}) \\
    Maximum number of single spin updates per submatrix update.

    \item \texttt{"neglect-Bennett-updates"}: boolean (\texttt{false}) \\
    Neglect the two types of Bennett updates:
    \begin{enumerate}
        \item Removal of an interacting spin that has already been proposed for removal.
        \item Removal of a non-interacting spin that has been proposed for insertion.
    \end{enumerate}
    Turning on this option leads to a larger number of delayed spins ($\approx$\texttt{max-submatrix-size}) at the cost of a systematic error due to the violation of detailed balance.

    \item \texttt{"additional-time-measurements"}: boolean (\texttt{false}) \\
    Do additional time measurements.
\end{itemize}}

\item \emph{SS-CT-HYB} \\
{\footnotesize
Used if \texttt{SS-CT-HYB} is selected as the cluster solver. \\[5pt]
Group \texttt{"SS-CT-HYB"}:

\begin{itemize}[label=$\bullet$, leftmargin=*]
    \item \texttt{"self-energy-tail-cutoff"}: integer (\texttt{0}) \\
    Cut-off parameter for the imaginary frequency tail of the self-energy.

    \item \texttt{"steps-per-sweep"}: double (\texttt{0.5})

    \item \texttt{"shifts-per-sweep"}: double (\texttt{0.5}) \\
    The fraction of insert/removal steps is given by \texttt{steps-per-sweep}/(\texttt{steps-per-sweep} + \texttt{shifts-per-sweep}), while the fraction of segments shifts is \texttt{shifts-per-sweep}/(\texttt{steps-per-sweep} + \texttt{shifts-per-sweep}).
\end{itemize}}

\end{itemize}

\subsection*{Four-point parameters}
{\footnotesize
Group \texttt{"four-point"}:

\begin{itemize}
    \item \texttt{"type"}: string (\texttt{"NONE"}) \\
    Four-point type, options are:
    \begin{itemize}
        \item \texttt{"NONE"}
        \item \texttt{"PARTICLE\_PARTICLE\_UP\_DOWN"}
        \item \texttt{"PARTICLE\_HOLE\_CHARGE"}
        \item \texttt{"PARTICLE\_HOLE\_MAGNETIC"}
        \item \texttt{"PARTICLE\_HOLE\_TRANSVERSE"}
    \end{itemize}

    \item \texttt{"momentum-transfer"}: array of doubles (null vector with the dimension of the lattice, e.g. for a 2D lattice: \texttt{[0., 0.]}) \\
    Transferred momentum $\mathbf{q}$.
    Must be an element of the reciprocal lattice of the DCA cluster ($\left\lVert \mathbf{q}-\mathbf{K}_\mathrm{DCA} \right\rVert_2 < 10^{-3}$).

    \item \texttt{"frequency-transfer"}: integer (\texttt{0}) \\
    Transferred frequency $\omega_n$.
    Given as the index $n$ of the bosonic Matsubara frequency $\omega_n = 2n\pi/\beta$.
\end{itemize}}

\subsection*{Analysis parameters}
{\footnotesize
Group \texttt{"analysis"}:

\begin{itemize}
    \item \texttt{"symmetrize-Gamma"}: boolean (\texttt{true}) \\
    Symmetrize the cluster and lattice irreducible vertex functions according to the symmetry group and diagrammatic symmetries.

    \item \texttt{"Gamma-deconvolution-cut-off"}: double (\texttt{0.5}) \\
    Cut-off parameter for the deconvolution of cluster irreducible vertex function.
    The deconvolution is done by Fourier transforming to real space, dividing by the coarse-graining patch functions in real space, $\phi(r)$, and then transforming back to reciprocal space.
    The real space patch functions are only inverted down to the value of this parameter,
    \begin{equation*}
    \phi^{-1}_\textrm{cut-off}(r) = \left\{\begin{array}{ll}
    \phi^{-1}(r), & \textrm{if} \,\, \phi^{-1}(r) > \texttt{Gamma-deconvolution-cut-off}, \\
    0, & \mathrm{else}.
    \end{array}\right.
    \end{equation*}

    \item \texttt{"project-onto-chrystal-harmonics"}: boolean (\texttt{false}) \\
    Project the product $\Gamma(k, k') \chi_0(k)$ onto the crystal harmonic functions and diagonalize in this subspace.

    \item \texttt{"projection-cut-off-radius"}: double (\texttt{1.5}) \\
    For the projection use crystal-harmonic functions of lattice vectors $\mathbf{r}$ with $\left\lVert \mathbf{r} \right\rVert_2 <$ \texttt{projection-cut-off-radius}.
\end{itemize}}

\subsection*{ED-solver parameters}
{\footnotesize
Only required if the exact diagonalization solver is used, e.g. in the application \texttt{cluster\_solver\_check}. \\[5pt]
\indent Group \texttt{"ED"}:

\begin{itemize}
    \item \texttt{"eigenvalue-cut-off"}: double (\texttt{1.e-6}) \\
    Only keep energy eigenvalues $E$ for which $e^{-\beta E} >$ \texttt{eigenvalue-cut-off}.
\end{itemize}}

\subsection*{Double-counting parameters}
{\footnotesize
Only required for LDA+DMFT. \\[5pt]
\indent Group \texttt{"double-counting"}:

\begin{itemize}
    \item \texttt{"method"}: string (\texttt{"none"}) \\
    The double-counting method to use, options are:
    \begin{itemize}
        \item \texttt{"none"}: no double-counting correction
        \item \texttt{"constant-correction-without-U-correction"}
        \item \texttt{"constant-correction-with-U-correction"}
    \end{itemize}

    \item \texttt{"correction"}: double (\texttt{0.}) \\
    The value of the double-counting correction.
\end{itemize}}

\newpage
\begin{lstlisting}[
basicstyle=\footnotesize\ttfamily,
caption={JSON-formatted sample input file for a DCA$^+$ calculation.},
frame=tb,
captionpos=t,
keywordstyle=\color{black},
label=listing:input_file]
{
    "output": {
        "directory": "./T=0.02/",
        "output-format": "HDF5",
        "filename-dca": "dca_tp.hdf5",
        "filename-analysis": "analysis.hdf5",
        "filename-profiling": "profiling.json",
        "dump-lattice-self-energy": false,
        "dump-cluster-Greens-functions": true,
        "dump-Gamma-lattice": false,
        "dump-chi-0-lattice": false
    },

    "physics": {
        "beta": 50,
        "density": 0.9,
        "chemical-potential": 0.,
        "adjust-chemical-potential": true
    },

    "single-band-Hubbard-model": {
        "t": 1.,
        "U": 4
    },

    "DCA": {
        "initial-self-energy": "./T=0.02/dca_sp.hdf5",
        "iterations": 1,
        "accuracy": 0.,
        "self-energy-mixing-factor": 1.,
        "interacting-orbitals": [0],

        "coarse-graining": {
            "k-mesh-recursion": 3,
            "periods": 2,
            "quadrature-rule": 1,
            "threads": 1,
            "tail-frequencies": 0
        },

        "DCA+": {
            "do-DCA+": true,
            "deconvolution-iterations": 16,
            "deconvolution-tolerance": 1.e-2
        }
    },

    "domains": {
        "real-space-grids": {
            "cluster": [[4, 6],
                        [8, -2]],
            "sp-host": [[20, 20],
                        [20,-20]],
            "tp-host": [[8, 8],
                        [8,-8]]
        },

        "imaginary-time": {
            "sp-time-intervals": 1024
        },

        "imaginary-frequency": {
            "sp-fermionic-frequencies": 1024,
            "four-point-fermionic-frequencies": 16
        }
    },

    "Monte-Carlo-integration": {
        "seed": 985456376,
        "warm-up-sweeps": 100,
        "sweeps-per-measurement": 1,
        "measurements-per-process-and-accumulator": 100,

        "threaded-solver": {
            "walkers": 2,
            "accumulators": 20
        }
    },

    "CT-AUX": {
        "expansion-parameter-K": 1.,
        "initial-configuration-size": 64,
        "initial-matrix-size": 128,
        "max-submatrix-size": 256,
        "neglect-Bennett-updates": false,
        "additional-time-measurements": false
    },

    "four-point": {
        "type": "PARTICLE_PARTICLE_UP_DOWN",
        "momentum-transfer": [0., 0.],
        "frequency-transfer": 0
    },

    "analysis": {
        "symmetrize-Gamma": true,
        "Gamma-deconvolution-cut-off": 0.5,
        "project-onto-crystal-harmonics": false
    }
}
\end{lstlisting}



\bibliographystyle{elsarticle-num}
\bibliography{refs.bib}







\end{document}